\documentclass{jair}

\setcopyright{cc}
\copyrightyear{2025}
\acmDOI{10.1613/jair.1.xxxxx}

\JAIRAE{Insert JAIR AE Name}
\JAIRTrack{} 
\acmVolume{4}
\acmArticle{6}
\acmMonth{8}
\acmYear{2025}

\RequirePackage[
  datamodel=acmdatamodel,
  style=acmauthoryear,
  backend=biber,
  giveninits=true,
  uniquename=init
  ]{biblatex}

\usepackage{enumerate}

\usepackage{amsmath,amsfonts}
\usepackage{textcomp}%
\usepackage{algorithm}%
\usepackage{algpseudocode}%

\usepackage{array}
\usepackage[caption=false,font=footnotesize]{subfig}
\usepackage{stfloats}
\usepackage{url}
\usepackage{verbatim}

\usepackage{hyperref}
\usepackage{enumerate}
\usepackage{tikz} 
\usetikzlibrary{positioning}
\usetikzlibrary{arrows.meta}
\usetikzlibrary{fit}
\usetikzlibrary{shapes.geometric}
\tikzset{
  plate/.style={draw=lightgray, shape=rectangle, rounded corners=0.5ex, thick,
    minimum width=1.1cm, text width=1.1cm, align=right, inner sep=10pt, 
	inner ysep=10pt,label={[xshift=0pt,yshift=10pt]south:#1}}
}

\theoremstyle{thmstyleone}%
\newtheorem{theorem}{Theorem}[section]
\newtheorem{lemma}{Lemma}[section]%
\theoremstyle{thmstyletwo}%
\newtheorem{remk}{Remark}[section]%
\theoremstyle{thmstylethree}%
\newtheorem{defn}{Definition}

\algnewcommand{\IfThenElse}[3]{
  \State \algorithmicif\ #1\ \algorithmicthen\ #2\ \algorithmicelse\ #3}
\algnewcommand{\IfThen}[2]{
  \State \algorithmicif\ #1\ \algorithmicthen\ #2\ }

\DeclareMathOperator*{\argmin}{arg\,min}

\usepackage{nameref}     

\begin{document}
\title[DCMAP]{Optimal Clustering with Dependent Costs in Bayesian Networks}

\author{Paul Pao-Yen Wu}
\authornote{Corresponding Author.}
\orcid{ 0000-0001-5960-8203}
\email{p.wu@qut.edu.au}
\affiliation{%
  \institution{Centre for Data Science, Queensland University of Technology}
  \city{Brisbane}
  \state{Queensland}
  \country{Australia}
}

\author{Fabrizio Ruggeri}
\orcid{0000-0003-0615-4417}
\email{fabrizio@mi.imati.cnr.it }
\affiliation{%
  \institution{Institute of Applied Mathematics and Information Technology, Italian National Research Council}
  \city{Milano}
  \state{}
  \country{Italy}}

\author{Kerrie Mengersen}
\orcid{0000-0001-8625-9168}
\email{k.mengersen@qut.edu.au}
\affiliation{%
  \institution{Centre for Data Science, Queensland University of Technology}
  \city{Brisbane}
  \state{Queensland}
  \country{Australia}
}

\renewcommand{\shortauthors}{Wu, Ruggeri \& Mengersen}

\begin{abstract}
{\bf Background:} 
Clustering of nodes in Bayesian Networks (BNs) and related graphical models such as Dynamic BNs (DBNs) has been demonstrated to enhance computational efficiency and improve model learning. It typically involves partitioning the underlying Directed Acyclic Graph (DAG) into cliques or optimising for some cost or criteria. 

    {\bf Objectives:}
We focus on a critical but understudied aspect of optimal clustering involving cost dependency. This is where inference outcomes and hence clustering costs depend on both nodes within a cluster and the mapping of clusters that are connected by at least one arc.

    {\bf Methods:}
 We propose a novel algorithm called Dependent Cluster MAPping (DCMAP) which can, given an arbitrary, positive cost function, iteratively and rapidly find near-optimal, then optimal cluster mappings. 
   
    {\bf Results:}
DCMAP is shown analytically to be optimal in terms of finding all of the least cost cluster mapping solutions and with no more iterations than an equally informed algorithm. Demonstrated on a complex systems seagrass DBN with $9.91\times10^9$ and $1.51\times10^{21}$ possible cluster mappings for 25 and 50 node configurations, it took 856 and 1569 iterations on average to find the first optimal solution, respectively.

    {\bf Conclusions:}
The effectiveness of DCMAP enables future research in BN learning using optimisation, such as through enhancing computational efficiency or minimising entropy for learning. This is critically important as computation of marginal distributions or updating model parameters is NP-hard. 

\end{abstract}



\received{10 December 2025}
\received[accepted]{x xx xxxx}

\maketitle

\section{Introduction}
\label{s:intro}
Bayesian Networks (BNs) and related methods including Dynamics BNs (DBNs) are a form of graphical model that are represented as Directed Acyclic Graphs (DAGs), where relationships between nodes (factors) are shown as arcs and quantified with conditional probabilities \citep{Pearl1988}. 
They have been applied widely across domains including medicine, environment, economics and social sciences to provide explanatory decision support, integrating data and even expert knowledge \citep{Wu2017,Kitson2023}. 
However, one of the key challenges associated with inference and learning of BNs and DBNs is computation, which is NP-hard \citep{Park2004,Murphy2002}.
To this end, different approaches to clustering of BN nodes have been used in the context of supporting more computationally efficient inference of marginal distributions given evidence, and model learning, to update the structure of the graph and/or the conditional probabilities associated with each node  \citep{Daly2011, Lauritzen1988, Kojima2010,Lu2021}. 
Appendix \ref{app:eg} illustrates with a simple example how three different clustering approaches reduce computational cost to different degrees.
A general approach to optimisation of BN clusters (i.e. with aribtrary criteria) can advance research in probabilistic graphical models by providing a benchmark for validating approximate inference methods, which are necessary in high-dimensional problems, and enable new methods for computation and inference such as by finding homogeneous or correlated regions in a non-homogeneous system \citep{Lin2020, Wu2018}.

Optimal clustering of a BN seeks to minimise some cost or objective function associated with the mapping of BN nodes to clusters. Typically, each node is uniquely mapped to a cluster and the total cost is the sum of costs associated with each cluster. However, a key challenge is cost dependency.
Consider a BN $H(\pmb{X},\pmb{A})$ with nodes $X_i\in\pmb{X}$ and arcs (or edges) $\pmb{A}$ that encode conditional dependence between parent and child nodes. Fundamentally, BN inference is underpinned by Bayes theorem where the joint probability distribution over all nodes $\pmb{X}$ and given observations (i.e.~evidence $\pmb{E}$) is \citep{Lauritzen1988}:
\begin{align}
P(\pmb{X},\pmb{E})&=\prod_{\forall X_j\in\pmb{X}}P(X_j|Par(X_j))\delta(X_j)
\label{sumprod}
\end{align}
where $P(X_j|Par(X_j))$ is the conditional probability of node $X_j$ given its parents, and $\delta(X_j)$ the evidence, which can be the inputs to or observations for inference. 
The joint probability distribution arises from the multiplication of conditional probability distributions, each of which is potentially multi-dimensional, comprising a node (or variable) $X_j$ and parent nodes $Par(X_j)$. Hence, intuitively: (i) clustering determines which conditional probability distributions are computed together, impacting computational complexity, and (ii) the computed outcomes of a given cluster are dependent on clusters connected by at least one arc via multiplication as per equation (\ref{sumprod}). 
This propagation of partial inference outcomes between clusters has been employed in approaches to BN and DBN inference such as \citet{Lauritzen1988} and \citet{Wu2018}. The dependency of the local cluster cost on the mapping of connected clusters is referred to as cost dependency, noting that the objective is to minimise the global cost which is a combination (typically additive) of local costs.
However, finding optimal clusters (i.e.~partitions) is NP-hard in general \citep{Buluc2016} and cost dependency exacerbates this complexity.

The clustering problem can be formalised as one of assigning a unique integer cluster label $u(X_j) =k, k=1,\ldots,m$ to each node $X_j\in\pmb{X}$ of a BN with DAG $H(\pmb{X},\pmb{A})$ and conditional probability parameters $\pmb{\theta}=\{P(X_j|Par(X_j)\}$; the total number of clusters $m$ is determined through optimisation. Since it is the DAG nodes that are clustered, for notational simplicity, let $G\bigl(H(\pmb{X},\pmb{A}),u(\pmb{X})\bigr)$ be the total cost to be minimised through cluster mapping $u(\pmb{X})$, implicitly capturing $\pmb{\theta}$ through the DAG. Additionally, $G^\ast\bigl(H(\pmb{X},\pmb{A})\bigr)$ is the optimal cost, and assume that the total cost is the sum of individual cluster costs $G_k$:
\begin{equation}
G^\ast\bigl(H(\pmb{X},\pmb{A})\bigr) = \argmin_{u(\pmb{X})}\Bigl( G\bigl(H(\pmb{X},\pmb{A}),u(\pmb{X})\bigr) \Bigr)  = \argmin_{u(\pmb{X})}\left( \sum_{k=1}^m{G_k} \right).
\label{argmineqn}
\end{equation}
Note that there may be multiple cluster mapping solutions $u^\ast_s(\pmb{X})$ with cost $G^\ast$; cost dependency implies that $G_k$ depends on the mapping of connected clusters.

\subsection{Case Study}
We use as a case study a DBN of a seagrass ecosystem \citep{Wu2018} to show how clustering can minimise computational cost. Seagrasses are a critical primary habitat for fish and many endangered species including the dugong and green turtle, and contribute USD\$1.9 trillion per annum through nutrient uptake and cycling. However, better management of human stressors such as dredging requires the modelling of cumulative effects arising from interactions between biological, ecological and environmental factors. 
The non-homogeneous DBN itself is complex and computationally intensive, and inference is infeasible within a standard (MCMC) framework for real-world decision support contexts.
Furthermore, understanding areas of uncertainty in the system and designing data collection strategies to mitigate uncertainty is of utmost interest in managing such complex systems. This uncertainty can be characterised by entropy with clusters corresponding to scenarios of evidence (i.e.~observed data) (Section \ref{ss:entropy}). 
We focus on computation as a cost function $G$ to demonstrate how the proposed algorithm can find clusters to minimise the computational complexity of inference, and discuss how our algorithm could also be used to minimise other cost functions such as entropy. 

\subsection{Existing Work}
\label{ss:existingwork}
Clustering of BNs has been employed in a wide range of contexts, evolving from a focus on computation of marginal distributions given evidence such as the seminal work of \citet{Lauritzen1988} to predominantly causal discovery and graph learning \citep{Kitson2023, Lu2021} in recent times. 
Almost all existing methods for inferring marginal distributions use heuristics based on the graph to form clusters, such as triangulation in join-tree inference \citep{Lauritzen1988}, conditioning sets (cutsets) in bucket elimination \citep{Dechter1999}, factor graphs \citep{Kschischang2001} and approximate inference with cutsets \citep{Bidyuk2007}. They help tackle the challenge of exponential growth in computational complexity with induced tree width \citep{Lin2020}. In some methods, such as mini-clustering \citep{Mateescu2010}, incremental region selection \citep{Forouzan2015}, clusters for selective filtering in a DBN \citep{Albrecht2016}, and triplet region construction \citep{Lin2020}, the heuristically formed clusters are refined through greedy search or local optimisation based on some function such as conditional entropy.
Much of the more recent work focuses on approximate methods to tackle high-dimensional problems. 
However, explicit study of global optimisation of clusters and their impact on computation and inference is missing in this literature. 

In contrast, the BN learning literature has seen some progress in clustering and inference towards global optimisation. There are three main approaches: (i) constraint based, that evaluate conditional independence structures i.e.~if an arc exists or not, then its direction, (ii) scoring, often with heuristics to evaluate goodness of fit of a proposed network, and (iii) hybrid, combining constraints with heuristics \citep{Huang2022} .

A number of approaches use clusters as a structure to restrict or constrain the local search space in a graph learning framework such as the seminal work of \citet{Friedman1999}. They define clusters in a similar vein to clique-trees \citep{Lauritzen1988} in a ``restrict and maximise'' framework to optimise a mutual-information based score of the learned network. \citet{Kojima2010} advance this approach, effectively relaxing the cluster tree requirement in \citet{Friedman1999} to a cluster graph to help address the issue of large clusters, which make local optimisation computationally infeasible. They used ancestral constraints to help limit the search space of clusters and the algorithm was shown to outperform greedy algorithms including Max-Min Hill Climbing (MMHC) and Hill Climbing (HC) with Tabu search. However, computational feasibility becomes a problem again when there are too many edges between clusters \citep{Kojima2010}. 

\citet{Lu2021} also adapt the work of \citet{Friedman1999} for undirected graphs, using clusters based on proposed parents and children of a node $X$ to constrain the search in an iterative framework based on A*, a heuristic-informed search algorithm. They show empirically that for small networks, exhaustive and in some cases optimal methods such as A* and SAT (Boolean SATisfiability) outperform greedy methods such as PC (Peter and Clark), which can suffer from convergence to local optimal, and their adaptation of A* performs as well as greedy methods in larger networks. However, their approach is exponential in complexity with neighbourhood size (i.e.~number of parents). 

\citet{Zhang2018} present an alternate approach in which clusters are determined through agglomerative clustering using Pearson correlation as the distance metric; however, little is provided in terms of algorithm performance and its convergence characteristics. \citet{Huang2022} and \citet{Gu2020} both develop comprehensive approaches that use similar agglomerative clustering based on mutual information and correlation as the distance metric, respectively. The former learns skeletons (an undirected version of the graph) for each cluster using their variant of the PC algorithm and log-likelihood ratio tests of conditional independence. \citet{Gu2020} use a score-based algorithm to learn partial DAGs for each cluster, which are then fused in the final step proposing edges between clusters. \citet{Huang2022} combine clusters with automated parameter tuning for turning skeletons into partial then fully directed acyclic graphs and hybrid greedy initialisation, obtaining empirical improvements in computational efficiency and more accurate DAGs. However, in both cases, the performance is dependent on the quality and size of the clusters. 

Clearly, clustering has demonstrated utility to improve computational feasibility in BN inference of marginal distributions and learning BNs with varying degrees of optimality given a scoring function (i.e.~cost function). 
However, none of the reviewed literature explicitly studied globally optimal clustering and its impact on inference and computation. Existing BN learning methods focus on inclusion or exclusion of edges, rather than scores that change given different mappings of nodes to clusters. 
Given the pervasive role clustering plays in BN research and the utility of dynamic programming and clustering for optimisation \citep{Friedman1999, Kojima2010, Lu2021}, this paper focuses on explicit, generalisable optimisation of clusters in the context of BN inference. 

Note that there also exists substantial literature in graph partitioning such as repeated graph bisection, Kernighan and Lin algorithm, spectral bisection and multilevel partitioning \citep{Buluc2016,Bader2013}. \citet{Zheng2020}, for instance, demonstrate how dynamic programming can be used to optimally and efficiently cluster nodes in the absence of cost dependency, applied to convolutional neural networks. However, the problem context for these approaches typically involves partitioning a graph into a set number of clusters or to satisfy some criteria for each cluster defined with respect to static node and/or edge weights. This differs significantly from the BN context and associated cost dependency between clusters. 

\subsection{Layout of Paper}
We develop a novel algorithm called DCMAP (Dependent Cluster MAPping) that optimises BN cluster mappings given an arbitrary, non-negative scoring function which we refer to as a cost function in the presence of cost dependency. Such dependency arises in the BN context of using clustering to improve computational efficiency when inferring marginal distributions given evidence \citep{Lauritzen1988} and in model learning \citep{Friedman1999,Lu2021}. 
A simple example is provided in Appendix \ref{app:eg} showing how DCMAP (least computation) and two other popular clustering approaches reduce computation. 
Noting as background the context of BN inference (Section \ref{ss:bninf}) and introducing an illustrative example (Section \ref{ss:eg}), we first present an intuitive overview of DCMAP and its notation (Section \ref{ss:overviewn}), followed by its derivation from dynamic programming principles (Section \ref{sss:layers} and \ref{sss:dp}). The algorithm is presented in detail in Section \ref{s:alg} and shown analytically to find all the optimal solutions such that an equally informed algorithm cannot do so in fewer iterations. An empirical demonstration is provided on the case study usage scenario of using clusters to minimise computation cost of inference for our complex systems seagrass DBN in Section \ref{s:sim}. Reflections and suggestions for future work, including extension to other cost functions for optimisation such as entropy are provided in the Discussion (Section \ref{s:discussion}).

\section{DCMAP Algorithm: Background and Overview}
\label{s:dcmap}
This section develops the foundational concepts for the proposed algorithm DCMAP and their relationship to BNs. It starts with some background on BN inference and an illustrative example of how clustering affects computation, provides an overview of the notation, and develops foundational concepts of layers, clusters and dynamic programming. Using these, a high-level conceptual presentation of the algorithm is provided along with the derivation of the objective function. Finally, an estimate of the size of the search space is provided to contextualise the complexity of cluster mapping.

\subsection{Background: BN Inference}
\label{ss:bninf}
In almost every usage scenario involving BNs including evaluation of what-if scenarios and model learning, it is necessary to compute marginal distributions for each node $X$ given evidence $P(X|\pmb{E})$ in the DAG $H(\pmb{X},\pmb{A})$. 
This marginal distribution is obtained from (\ref{sumprod}) by marginalising (i.e.~summing) out all other nodes:
\begin{align}
P(X|\pmb{E})&=\frac{P(X,\pmb{E})}{P(\pmb{E})}\nonumber\\
P(X,\pmb{E})&=\sum_{\forall X_i\in\pmb{X}\backslash X}\prod_{\forall X_i\in\pmb{X}}P(X_i|Par(X_i))\delta(X_i) 
\label{sumprod1}
\end{align}
where $P(X_i|Par(X_i))$ is the conditional probability of node $X_i$ given its parents, $\backslash$ denotes set difference, and $\delta(X_i)$ is the evidence for node $X_i$. 
Evidence can be used to, among other things, incorporate observations and evaluate scenarios.
The joint distribution (\ref{sumprod}) has $|\pmb{X}|$ nodes or dimensions, and $\sum_{X_i}$ denotes a summation over all discrete states of node $X_i$ which we denote as lowercase $x_{i1},x_{i2},\ldots$, effectively removing that dimension from the joint distribution through marginalisation.
Note that $P(\pmb{E})$ is a constant that normalises the resulting distribution \citep{Lauritzen1988}. 
Although the notation focuses on discrete probability distributions to minimise complexity, the proposed clustering algorithm is generalisable to continuous distributions since it ultimately relies only on an arbitrary non-negative cost function.

\subsection{Illustrative Example}
\label{ss:eg}
Intuitively, if the clustering cost function to be optimised (\ref{argmineqn}) is dependent in some way on BN inference (\ref{sumprod1}), then it will be affected by cost dependency due to the propagation of probabilities via parent-child arcs between clusters. To help illustrate this and the proposed algorithm, consider an example discrete BN DAG (Fig. \ref{fig:eg}).
Assuming binary states for all nodes, the Conditional Probability Table (CPT) for each node with no parents (nodes $A$, $B$, $C$) is a $2\times 1$ matrix, e.g.~$P(A)=[p_a,p_{\bar{a}}]^T$ for states $A=a$ and $A=\bar{a}$. The CPT for $F$ is a $2\times 2$ matrix, i.e.~$P(F|A)=\begin{bmatrix} p_{af} &p_{\bar{a}f}\\ p_{a\bar{f}} & p_{\bar{a}\bar{f}}\end{bmatrix}$ where $p_{af}$ is the probability of state $F=f$ given state $A=a$. $P(D|A,B)$ and $P(G|D,E)$ are $2\times 4$ matrices. Multiplication of CPTs (\ref{sumprod}) containing different nodes, i.e.~variables, grows the dimensionality of the resultant matrix with associated exponential growth in computation; in the discrete case, the resultant matrix contains all possible combinations of node states. We ignore evidence terms for notational simplicity, noting that evidence can be easily incorporated using $\delta(X_i)$ with node $X_i$ (\ref{sumprod}). Using (\ref{sumprod1}), the posterior probability $P(F)$ is:
\begin{align}
\footnotesize
P(F) =& \sum\limits_{A,B,C,D,E,G}P(A)P(B)P(C)P(D|A,B)P(E|C)P(F|A)P(G|D,E)\label{egjoint}\\
=& \sum\limits_A P(F|A)P(A) \left(\sum\limits_{D,E,G} P(G|D,E) \left(\sum\limits_B P(D|A,B)P(B)\left( \sum\limits_C P(E|C)P(C)\right)\right)\right)\label{egcluster}
\end{align}
Consider, as an example, a cost function based entirely on computational cost, and one approach where all nodes are in one cluster (\ref{egjoint}) where the product of conditional probabilities produces a matrix of $7$ dimensions, one for each node $A$ through $G$, with exponential growth in memory ($2^7$ joint probability values) and computation associated with multiplication ($608$ products) and marginalisation ($126$ sums). For example, $P(A)P(B)$ requires $4$ multiplications, and $P(A)P(B)P(D|A,B)$ requires $4+8=12$. $\sum_{A}P(F|A)$ requires $2$ summations. 

Contrast this with the cluster mapping shown through shading (e.g.~$A$ and $F$ are in the same cluster, as are $D$ and $G$; all others in a cluster of their own) in Fig. \ref{fig:eg}. We illustrate clustering with a simple approach where probabilities from an antecedent cluster are multiplied with those in the current cluster then marginalised using an arbitrary ordering. The resultant computation (\ref{egcluster}) has maximum dimensionality of $4$ due to interspersed multiplication and marginalisation; recall marginalisation happened at the end in (\ref{egjoint}). In this example, clustering greatly reduces computation and memory cost. It also demonstrates cost dependency, since the computation cost required at a descendant cluster is dependent on the mapping and outcomes of antecedent clusters. Finally, it demonstrates the forwards and backwards nature of inference, such as forwards propagation from cluster $E$ to cluster $\{D,G\}$, and backwards propagation from $\{D,G\}$ to cluster $\{A,F\}$. For the full example, see Appendix \ref{app:eg}.
\begin{figure}
\centering
\footnotesize
	\begin{tikzpicture}[>={Latex[width=2mm,length=3mm]},
						every node/.style={circle,thick,draw},
						every edge/.style={draw=black,thick}]
		\node (F) [] at (0,0) {F};
		\node (G) [fill=black,text=white, below =of F]  {G};
		\node (D) [fill=black,text=white, left =of G] {D};
		\node (E) [fill=gray!85, below =of D] {E};
		\node (B) [fill=gray!25, left =of D] {B};
		\node (A) [ above =of B] {A};
		\node (C) [fill=gray!60, below =of B] {C};
	
		\path [->] (D) edge (G);
		\path [->] (E) edge (G);
		\path [->] (A) edge (F);
		\path [->] (A) edge (D);
		\path [->] (B) edge (D);
		\path [->] (C) edge (E);
	
		\node[plate=Layer 0, inner sep=10pt, fit=(F) (G) ] (plate0) {};
		\node[plate=Layer 1, inner sep=10pt, fit=(D) (E) ] (plate1) {};
		\node[plate=Layer 2, inner sep=10pt, fit=(A) (B) (C)] (plate2) {};
	
	\end{tikzpicture}
\caption{Example DAG showing nodes, directed edges, and layers where shading denotes clusters obtained using DCMAP. Assume all nodes have binary states.}
\label{fig:eg}
\end{figure}

\subsection{Overview and Notation}
\label{ss:overviewn}
In short, given a BN $H(\pmb{X},\pmb{A})$, the task is to find one or more cluster mappings $U_s(\pmb{X})$ that minimise the total cost as per equation (\ref{argmineqn}). 
We first define layers (Section \ref{sss:layers}) and clusters (Section \ref{sss:clusters}) as a way to decompose the DAG to enable step-by-step optimisation via dynamic programming (Section \ref{sss:dp}). Notation is summarised in Table \ref{tab:defn}. 
\begin{table}
\footnotesize
\caption{Summary of notation. Generally, we use lower case notation (e.g.~$g,u$) for local calculation, upper case (e.g.~$G,U$) for global search and optimisation, non-italicised (e.g.~$\text{G},\text{U}$) for data structures in our algorithm, and \textbf{bold} for a set. We notate $k=1..n$ to denote $k=1,2,...,n$ for brevity.}
\begin{tabular}{|p{0.12\columnwidth}|p{0.8\columnwidth}|}
\hline\hline
Symbol	& Definition	\\
\hline
$H(\pmb{X},\pmb{A})$, $n$, $\pmb{X}^\textrm{leaf}$	& DAG $H$ with $n$ nodes $X\in\pmb{X}$, leaf nodes $\pmb{X}^\textrm{leaf}\in\pmb{X}$, and arcs $A$.\\
$Par()$, $Child()$ &These functions return the parent and child nodes / cluster / cluster-layers, respectively, of a given node(s) / cluster(s) / cluster-layer(s).\\
$G()$		& Objective function cost given nodes and cluster mappings; $G^\ast$ is the optimal cost (Section \ref{sss:dp}) and $\text{G}_\text{min}$ is the current minimum global cost in the search. \\
$c()$		& Transition cost for one step of dynamic programming (\ref{dpeqng}); positive, non-zero function of nodes and associated conditional probabilities in a cluster-layer and child cluster-layer(s).\\
$l$, $\pmb{X}_l$			& Layer, $l=0$ (i.e.~leaf nodes), last layer $l_\text{max}$; $L$ is the largest i.e.~latest layer of a parent node (Section \ref{sss:layers}). $\pmb{X}_l$ denotes nodes in layer $l$. \\
$k$, $\pmb{X}_k$, $\phi(\pmb{X}_k)$, $\pi(\pmb{X}_k)$			& Cluster label $k$ with $\pmb{X}_k$ denoting nodes in that cluster, $\phi(\pmb{X}_k)$ and $\pi(\pmb{X}_k)$	internal and link nodes (Section \ref{sss:layers}). \\
$kl$, $\pmb{X}_{kl}$	&Cluster-layer label and set of cluster-layer nodes, respectively (Section \ref{sss:dp}).\\
$u(X), U$		& Cluster mapping of node $X$ to integer cluster label $k$; $U$ is a map for a vector of nodes, $\hat{\textrm{U}}$ a mapping proposal, and $U^\ast$ a mapping with optimal cost $G^\ast$. \\
$J()$		& Function to capture dependencies between a cluster-layer and its child cluster-layer(s) (Section \ref{sss:dp}).\\
$\hat{g}$	& Heuristic estimate of the total cost (Section \ref{sss:dp}).\\
$h()$		& Heuristic estimate of true cost $G()$ given remaining nodes; used to compute $\hat{g}$ above.\\
$Q$			& Queue for prioritising the search; $Q'$ are eligible queue entries for popping (Section \ref{s:alg}).\\
$b$, $\text{Br}_\text{lk}$, $\text{Br}_\text{pl}$			&Branch $b$ of the search and data structures for branch linking $\text{Br}_\text{lk}$ and branch updates $\text{Br}_\text{pl}$ (Section \ref{s:alg}).\\
\hline\hline
\end{tabular}
\label{tab:defn}
\end{table}

\subsubsection{Layers}
\label{sss:layers}
\begin{defn}
Given a DAG $H(\pmb{X},\pmb{A})$, define the layer $l(X_j)=l$ of node $X_j\in\pmb{X}$ as the length of the longest path (i.e.~number of arcs) following the direction of the arcs (parent to child) to the set of leaf nodes $X\in \pmb{X}^\textrm{leaf}$, i.e.,~nodes with no child nodes. 
\label{defn:layer}
\end{defn}
We assume that $H$ is fully connected, since cluster mapping could be applied independently to each connected component. 
We use a simple breadth-first search \citep{Lavalle2006} to compute $l(\pmb{X})$ in $O(n)$ time (Algorithm \ref{alg:layer}).
\begin{algorithm}
\footnotesize
\caption{Layer assignment algorithm; $Par(\pmb{X}')$ denotes the union of all parent nodes of a set of nodes $\pmb{X}'$.}
\label{alg:layer}
\begin{algorithmic}[1]
\Procedure{Layer}{}
\State Initialise $l\gets 0$, $\pmb{X}' \gets \pmb{X}^\textrm{leaf}$, $l(\pmb{X}) \gets \infty$
\State $l(\pmb{X}') \gets 0$
\While{$Par(\pmb{X}')\neq \varnothing$}
\State $\pmb{X}' \gets Par(\pmb{X}')$
\State $l \gets l+1$
\State $l(\pmb{X}') \gets \max\left(l(\pmb{X}'),l\right)  $
\EndWhile
\EndProcedure
\end{algorithmic}
\end{algorithm}
\begin{lemma}
\label{lemma:layer}
Any pair of nodes in the same layer $(X_1,X_2): l(X_1)=l(X_2)=l$ must be disjoint, i.e.~there cannot be an arc between them. 
\end{lemma}
This can be easily shown through contradiction since if there were an arc from $X_1$ to $X_2$, then $X_1$ is a parent of $X_2$ and Algorithm \ref{alg:layer} would have updated $l(X_1)$ to $l+1$ at the next iteration on line 7, which means they cannot be from the same layer. Hence, all nodes in the same layer are disjoint (see Fig. \ref{fig:eg} for an example). Additionally, since there are no cycles in a DAG, Algorithm \ref{alg:layer} is guaranteed to terminate after at most $n=|\pmb{X}|$ iterations. 
\begin{remk}
\label{remk:layer}
As the layer $l(X_1)$ is by definition the longest path to a leaf node, then $l(X_1)\geq l(X_2)+1$ where $X_2$ is a child of $X_1$ (equivalently $X_1 \in Par(X_2)$); potentially, there exists a longer path from $X_1$ to a leaf node than $l(X_2)+1$.
\end{remk}

\subsubsection{Clusters}
\label{sss:clusters}
\begin{defn}
\label{defn:k}
Nodes are mapped to clusters using integer labels $u(X)=k$ where $1\leq u(X) \leq m$ for $m$ unique clusters, $m\leq n$. Denote the set of nodes in cluster $k$ as $\pmb{X}_k$. Denote $u'(X)$ as a proposed cluster mapping for node $X$. 
\end{defn}
Clusters must be contiguous (Definition \ref{defn:kcontig}) to avoid possible introduction of cycles between clusters. 
\begin{defn}
\label{defn:kcontig}
A cluster $\pmb{X}_k$ is contiguous iff given any two nodes $X_1,X_2\in\pmb{X}_k$, there are no parent-child paths $\pmb{X}_p=\{X_1,\ldots,X',\ldots,X_2\}$ with a node $X'$ mapped to a different cluster.
\end{defn}
BN inference assumes a DAG with no cycles to ensure convergence when propagating evidence \citep{Pearl1988}. This extends to clusters (or cliques) to enable computationally efficient propagation of evidence between clusters as part of BN inference \citep{Lauritzen1988}.
As an example, if in Fig. \ref{fig:eg} nodes $B,G$ were in cluster $1$ and $D$ in cluster $2$, then that breaks the contiguity of cluster $1$ and introduces a cycle between cluster $1$ and $2$ via the path $\{B,D,G\}$. If instead nodes $D,G$ were in cluster $1$ and $B$ in cluster $2$, then contiguity is maintained and there are no cycles between cluster $1$ and $2$.
Since the underlying DAG is acyclic, it is easy to see that, with contiguity, the resulting clusters are also acyclic.
\begin{defn}
Define a cluster-layer $kl$ as a subset of nodes $\pmb{X}_{kl}\subseteq \pmb{X}_k$ comprising all the nodes in layer $l$ of cluster $\pmb{X}_k$. Therefore, $\bigcup_{l=l^0_k}^{l^1_k}\pmb{X}_{kl} = \pmb{X}_k$ where $l^0_k$ and $l^1_k$ are the earliest and latest layers of cluster $\pmb{X}_k$; hence $l^1_k\geq l^0_k$.
\label{defn:clustlayer}
\end{defn}

\begin{defn}
\label{defn:k-lkint}
Nodes in cluster $\pmb{X}_k$ with at least one child node in a different cluster are denoted as link nodes $\pi(\pmb{X}_k)$, and the remaining nodes are internal nodes $\phi(\pmb{X}_k)$, i.e.~$\pmb{X}_k \backslash \pi(\pmb{X}_k) = \phi(\pmb{X}_k)$.
\end{defn}
In a similar vein, parent $Par(\pmb{X}_k)$ or child clusters $Child(\pmb{X}_k)$ must have at least one arc linking them to cluster $\pmb{X}_k$, and parent $Par(\pmb{X}_{kl})$ or child cluster-layers $Child(\pmb{X}_{kl})$ also have at least one arc linking them to $\pmb{X}_{kl}$. 
Note, a cluster mapping $u(\pmb{X}_k)$ induces a unique mapping of link and internal nodes based on parent-child relationships between different clusters. However, the reverse is not necessarily true. Hence, our optimisation task focuses on finding $u(\pmb{X}_k)$. 

\subsubsection{Dynamic Programming}
\label{sss:dp}
We use dynamic programming to find optimal cluster mappings.
Generally, dynamic programming is formulated as the optimisation of an objective function (or cost function) $G$ over a series of decisions or actions $u_k$ for $k=0,\ldots,K-1$ where the optimal cost $G^\ast$ over an interval $[k,K]$ is \citep{Bellman1962}:
\begin{equation}
G_{(k,K)}^\ast(x) = \inf_{u_k\in \mathcal{A}_k}\left[G_{(k,K)}\left(x_k,u_k,U_{(k+1,K-1)}^\ast\right)\right]
\label{dpeqn}
\end{equation}
where $U_{(k+1,K-1)}^\ast$ is the optimal sequence of actions $u_{k+1},\ldots,u_{K-1}$ for the interval $[k+1,K-1]$, $\mathcal{A}_k$ is the set of possible actions at stage $k$, $x$ is the state at stage $K$ as a consequence of optimal choice of $u_k$ at state $x_k$, and $\inf$ is the infimum operator which returns the argument that gives the minimum expression. 
Intuitively, given the current state $x_k$ and an optimal sequence of actions at stages $k+1$ to $K-1$ leading to state $x=x_K$, we choose the action $u_k$ to take at the current stage to minimise the overall cost in going from stage $k$ to $K$. 
This formulation is the most flexible as the cost function can vary (i.e~time-varying in the literature, but we refer to as stage varying to avoid confusion with time in a DBN) across intervals $[k,K]$, and does not have to be additive. 
This is useful as there can be multiple clusters in a given layer of the DAG to cluster (Section \ref{sss:highlevel}).
However, computational complexity grows in NP time with graph size and with the number of proposals $|\mathcal{A}_k|$ at each stage \citep{Lavalle2006}.  

If the objective function $G$ is additive across stages and independent of $k$ (implying that the system is time invariant), then (\ref{dpeqn}) simplifies to
\begin{equation}
G^\ast_{k+1}(x) = \inf_{u\in \mathcal{A}} \left[c(x,u) + G_k^\ast\left(f(\pmb{x},\pmb{u})\right) \right]
\label{dpeqnsimple}
\end{equation}
where $f(\pmb{x},\pmb{u})$ is a sequence of states and actions, starting at stage $k=0$, resulting in state $x$ at stage $k$ with optimal cost $G^\ast_k$, and $c$ is the transition cost of enacting action $u$ at state $x$. 
An additive and time-invariant system greatly reduces computational complexity.
Optimisation algorithms are built around the objective function, and Equation (\ref{dpeqnsimple}) is the basis for many heuristic search techniques such as A*. A heuristic estimate of the remaining cost over stages $[k+1,K]$ is used in A* to prioritise the search and potentially reduce computation time, such as for BN learning \citep{Lu2021}. 

\subsection{High-Level Algorithm}
\label{sss:highlevel}
The cost function being optimised through clustering can be arbitrary, but its computation differs from inferring marginal distributions for every node (\ref{sumprod1}) as the latter requires forwards and backwards propagation, whilst the former computes a global cost based on propagation in one direction, akin to scoring in BN learning (Section \ref{ss:existingwork}). This cost function could be the computational cost associated with cluster-based inference (equation \ref{sumprod1}), entropy, predictive error, or information criterion such as the Watanabe-Akaike Information Criterion (WAIC).

The proposed algorithm DCMAP uses layers $l$ as dynamic programming stages in equation (\ref{dpeqn}) for incremental optimisation of the cost, beginning at layer $l=0$ (i.e.~leaf nodes) and working backwards through the BN from earlier to later layers, culminating at $l_\text{max}$. A queue $Q$ is used to order the evaluation of possible cluster-layer mappings $\hat{\textrm{U}}$, which are linked to connected cluster-layers, by lowest overall cost first.
The overall cost includes the previous layer cost $G(\pmb{X}_{l-1})$, the cost $c$ of the proposed cluster-layer, and a heuristic estimate $h$ of the remaining cost (\ref{dpeqng2}). 

Conceptually, different cluster-layer mappings could be thought of as branches in the search tree, which are linked at layer $l$ if they share the same mappings $u$ at layers $l=0,\ldots,l-1$. Intuitively, an efficient algorithm prunes branches that are dominated at the earliest opportunity to minimise computation time. These high level components of the algorithm are illustrated in Fig. \ref{fig:overall}.
\begin{figure}
\centering
\footnotesize
	\begin{tikzpicture}[>={Latex[width=2mm,length=3mm]},
						every node/.style={rectangle,rounded corners,minimum width=3cm,minimum height=0.8cm, thick,draw},
						every edge/.style={draw=black,thick},
						plate/.style={draw, label={[xshift=70pt,yshift=-6pt]north west:#1}} ]
		\node (1) [] at (0,0) {Initialisations (Section \ref{sss:dataandinit})};
		\node (2) [below=of 1] {\parbox{5cm}{Pop $kl$ from $Q'$, or activate a branch if no eligible $kl$ and try again (Section \ref{sss:pop})}};
		\node (3) [below=of 2] {\parbox{5cm}{Prune dominated branches (linked or global threshold $G_\text{min}$) (Section \ref{sss:prune})}};
		\node (4) [below=of 3] {\parbox{5cm}{Propose cluster mappings for $\pmb{X}_{kl}$ based on $\hat{\textrm{U}}$ (Section \ref{sss:proposalscl})}};
		\node (5) [below=of 4] {\parbox{5cm}{Compute costs and update branches (Section \ref{sss:compute}). Least cost branch is active.}};
		\node (6) [below=of 5] {\parbox{5cm}{If a layer is completely mapped (Section \ref{sss:layerupd}), update linked branches, update $G_\text{min}$ if $l=l_\text{max}$.}};
		\node (7) [below=of 6] {\parbox{5cm}{Update possible $kl$ mappings for this branch in $\hat{\textrm{U}}$ and put this $kl$ and branch into $Q$ (Section \ref{sss:updateUQ})}};
		\path [->] (1) edge (2);
		\path [->] (2) edge (3);
		\path [->] (3) edge (4);
		\path [->] (4) edge (5);
		\path [->] (5) edge (6);
		\path [->] (6) edge (7);

		\node[plate=For each $\pmb{X}_{kl}$ proposal, inner sep=10pt, fit= (5) (6) (7) ] (plate1) {};
		\node[plate=While $Q'$ is not empty, inner sep=10pt, fit=(2) (3) (4) (5) (6) (7) (plate1) ] (plate0) {};

%
%
	\end{tikzpicture}
\caption{Illustration of the high level components of the proposed algorithm DCMAP.}
\label{fig:overall}
\end{figure}

\subsection{Objective Function and Heuristic}
\label{ss:objectivefcn}
Cost dependency and the potential for the parent $Par(X)$ of a child node $X\in\pmb{X}_l$ to be at a layer $l'>l+1$ (Remark \ref{remk:layer}) violates the assumption of time invariance (\ref{dpeqnsimple}). Hence, a new objective function is derived that can accommodate this for the proposed cluster mapping algorithm (Fig. \ref{fig:overall}).

The structure of layers $\pmb{X}_l$ remain fixed whilst cluster-layers $\pmb{X}_{kl}$ change in a cluster-mapping setting. Hence, layers could be used as a structure for a time-invariant, objective function.
Let the latest antecedent layer of a parent be $L$:
\begin{equation}
L = \max\Bigl( l\bigl(Par(\pmb{X}_l)\bigr)\Bigr)
\label{parlayer}
\end{equation}
Substituting into the general objective function of \citet{Bellman1962} (\ref{dpeqn}), the objective function over layers $[l,L]$ is:
\begin{align}
G_{(l,L)}^\ast(x) &= \inf_{u(\pmb{X}_l)\in U_l}\left[G_{(l,L)}\left(\pmb{X}_l,u(\pmb{X}_l),U_{(l+1,L)}^\ast\right)\right]
\label{dpeqnL}
\end{align}
where $u$ describes a cluster mapping and $U^\ast$ a series of optimal cluster mappings. 
Here, the propagation of probabilities during inference occurs between nodes in layer $l$ and parent nodes in layers $l+1$ up to and including $L$, which means nodes at these layers need to be considered when finding the optimal cluster mappings at layer $l$. The objective function is thus stage-varying and, in this formulation, not necessarily additive. 
However, such a cost function is computationally expensive to evaluate. 

As disjoint nodes are conditionally independent and nodes in a layer or cluster-layer are disjoint by Definition \ref{defn:clustlayer}, we assume that the objective function cost of disjoint cluster-layers are independent and thus additive. 
Moreover, by definition of $L$ (\ref{parlayer}) as the latest parent layers given nodes at layer $l$, an additive and time-invariant formulation for the objective function could be obtained by finding the lowest cost cluster-mapping (the infimum) over all intermediary cluster-layers. This replaces the $U^*_{l+1,L}$ term in (\ref{dpeqn}) and makes $G$ time invariant. Thus, (\ref{dpeqnL}) can be re-written as:
\begin{align}
\footnotesize
&G^*(\pmb{X}_{L}) = G^*(\pmb{X}_{l}) + \nonumber\\
&\inf_{u(\pmb{X}_{l+1,\ldots,L})}\sum_{l'=l+1}^{L}{\sum_{k=1}^{m_{l'}}{c\Bigl( 
	\pmb{X}_{kl'},  u(\pmb{X}_{kl'}), Child(\pmb{X}_{kl'}), u\bigl(Child(\pmb{X}_{kl'})\bigr)\Bigr) }}
\label{dpeqng}
\end{align}
where the optimal cost $G^*(\pmb{X}_{L})$ is only a function of the optimal cost at layer $l$ and cluster-layer mappings between $l+1$ and $L$ inclusive, notated as $u(\pmb{X}_{l+1,\ldots,L})$, and $c$ is a positive, non-zero cost that is a function of nodes, cluster mappings and implicitly associated conditional probabilities in a cluster-layer and child cluster-layer(s). Note that $m_{l'}$ is the number of unique clusters at layer $l'$. In the BN inference context (Section \ref{ss:eg} and (\ref{sumprod1}), cost dependency arises from propagation between child cluster-layers $Child(\pmb{X}_{kl'})$, their cluster mapping, and a parent cluster-layer $kl'$ and its cluster mapping; potentially, they could be of the same cluster.

Equation (\ref{dpeqng}) provides the basis for our algorithm. However, in algorithms like A* \citep{Lavalle2006}, potential actions (i.e.~cluster mappings) are evaluated prospectively using the objective function at each iteration. This rapidly becomes infeasible in our cluster mapping context due to the combinatorial explosion of mappings over multiple layers $u(\pmb{X}_{l+1,\ldots,L})$. One approach to overcome this challenge is to instead apply our objective function (\ref{dpeqng}) retrospectively after potentially multiple iterations of the algorithm, to prune dominated branches (Section \ref{ss:theoretical} and Lemma \ref{lemma:G}).
\begin{defn}
A cluster mapping solution for branch $i$ on layer $l$ is dominated by another branch $j$ iff $G_i(\pmb{X}_l)>G_j(\pmb{X}_{l})$.
\label{defn:dominate}
\end{defn}

Here, it is necessary to have an alternate means for prioritising cluster-layer proposals to update at each iteration.
Two key heuristics are used: 
(a) an upper bound estimate of the total cost $\hat{g}$, and (b) adaptive constraint and release of the search space (Section \ref{sss:pop}). 

Recall that $l_\text{max}$ is the last layer of the DAG, hence $\hat{g}$ is an estimate of the total or global cost $G(\pmb{X}_{l_\text{max}})$, comprising the objective function cost for this branch $G(\pmb{X}_{l-1})$, the cost for clusters in layer $l$, and a heuristic estimate of the remaining cost:
\begin{align}
\hat{g}(\pmb{X}_{kl}) &= G(\pmb{X}_{l-1}) + \nonumber\\
&\sum_{k\in M_l}{c\Bigl( 
	\pmb{X}_{kl'},  u(\pmb{X}_{kl'}), Child(\pmb{X}_{kl'}), u\bigl(Child(\pmb{X}_{kl'})\bigr)\Bigr) } +\nonumber\\
&h\Bigl( \pmb{X} \setminus \bigl(\pmb{X}_{kl : k\in M_l}\cup\pmb{X}_{0..l-1}\bigr) \Bigr)
\label{dpeqng2}
\end{align}
where $M_l$ is a set of clusters at layer $l$ that have been updated, $\setminus$ is set difference, and $h$ is a heuristic cost that is an upper bound for all the remaining, non-updated nodes. We refer to nodes for which the cost has been calculated as updated or popped. Potentially, there may be some nodes remaining in layer $l$ that have not been updated which are estimated as part of the $h$ term (\ref{dpeqng2}). Note that it is only possible to update cluster-layers at layer $l$ when all nodes in layer $l-1$ and earlier layers have been updated at least once. 

\subsection{Search Space Size}
\label{ss:sss}
Using the concept of layers, we can approximate the size of the search space of our cluster mapping problem. Naively, a DAG with $n$ nodes could be mapped to $k=1..n$ clusters thus resulting in $n^n$ possible mappings. However, due to the requirement for cluster contiguity (Definition \ref{defn:kcontig}), there are many infeasible proposals. A simple approach to ensuring contiguity is to either map a node to a new cluster unique to itself, or to an existing cluster of a child node (Section \ref{sss:proposalscl}). The overall search space size $\eta$ arises from the combination of layer by layer cluster mapping combinations, i.e.,~the product of products (\ref{ssseqn}). 
\begin{equation}
\eta = \prod_{l=1..l_\text{max}}{\prod_{X_i\in\pmb{X}_l}{\bigl(\left|Child(X_i)\right|+1\bigr)}}
\label{ssseqn}
\end{equation}
As might be expected, the search space grows exponentially with layers, and the out-degree of the DAG can also significantly grow the size of the search space \citep{Buluc2016}. In- and out-degree refer to the number of parent or child nodes, respectively \citep{Buluc2016}. 
For instance, the case study DBN \citep{Wu2018}, with small world connectivity, has just 25 nodes but a search space of 9.9 billion mapping combinations. This study and corresponding DCMAP solution are described in Section \ref{s:sim}.

\section{Algorithm and Analysis}
\label{s:alg}
This section formalises the components of the algorithm in Fig. \ref{fig:overall}, and proves the optimality of the method.

\subsection{Algorithm Components}
Our method uses $\hat{g}(\pmb{X}_{kl})$ to prioritise cluster-layer proposals to update (i.e.~pop), and $G(\pmb{X}_l)$ to facilitate local optimisation by identifying and cutting off further exploration of dominated solutions. It iterates from layer $0$ (i.e.~leaf nodes) through to $l_\text{max}$, storing unique and non-dominated cluster mappings on separate search branches. The algorithm terminates if, given current cluster-layer proposals on the queue and additive costs, the current costs of these potential solutions exceed the least-cost solutions found so far. Potentially, near-optimal solutions can be returned at earlier iterations of the search.

\subsubsection{Data Structures and Initialisations}
\label{sss:dataandinit}
The main data structures are sparse matrices, which only store non-zero entries and assume zero otherwise for memory efficiency. 
The $j^{th}$ row (corresponding to branch $j$), $k^{th}$ column (cluster $k$) and $i^{th}$ (node $i$) slice of a three-dimensional matrix $\mathcal{M}$ is indexed by $\mathcal{M}[j,k,i]$.
The structures are defined in Table \ref{tab:struct}. Note the distinction between proposed versus realised cluster mappings with associated updated costs stored in $\hat{\textrm{U}}$ and $U$, respectively. $\text{Br}_\text{lk}$ is used for pruning linked branches that are dominated and $\text{Br}_\text{pl}$ for tracking (using layers) which queue entries on that branch are eligible to be popped.
\begin{table}
\footnotesize
\caption{Definition of data structures. We use $\$$ to denote a named field within a data structure; for example, the layer field $l$ within a popped queue entry $q$ is denoted $q\$l$.}
\begin{tabular}{|p{0.25\columnwidth}|p{0.7\columnwidth}|}
\hline\hline
Data Structure	& Definition	\\
\hline
$\hat{\textrm{U}}\in\mathbb{R}^{B\times n\times n}$	&Proposed cluster mappings before updates, where $\hat{\textrm{U}}[j,k,i]=1$ indicates a proposal of cluster $k$ for node $i$ on branch $j$.\\
$\textrm{U}\in\mathbb{R}^{B\times n}$				&Cluster mapping $\textrm{U}[j,i]=k$ after updates are made at each iteration.\\
$\begin{aligned}[t]\text{Br}_\text{lk}\in\nonumber\\\mathbb{R}^{B\times l_\text{max}\times B\times l_\text{max}}\nonumber\end{aligned}$		&$\text{Br}_\text{lk}[i,l,j,l]=1$ indicates a link at layer $l$ between branches $i$ and $j$.\\
$\text{Br}_\text{pl}\in\mathbb{R}^{B}$					&Stores $l^{*}+1$ where $l^{*}$ is the highest layer where all nodes have been updated at least once on branch $b$.\\
$\text{G}\in\mathbb{R}^{B\times l_\text{max}}$		&Current objective function costs $G(\pmb{X}_l)$ (\ref{dpeqng}) for each branch and layer.\\
$\text{G}_\text{min}$					&Current minimum total cost $G(\pmb{X}_{l_\text{max}})$.\\
$Q$, $Q'$								&A queue with entries $q=\{b,u(\pmb{X}_{kl}),l,\hat{g}(\pmb{X}_{kl}),G(\pmb{X}_l)\}$. We denote $Q'$ as eligible entries for updating (popping) $q\in Q:\text{Br}_\text{pl}[b]\geq q\$l$, aka activated branches.\\
\hline\hline
\end{tabular}
\label{tab:struct}
\end{table}

At initialisation, the objective functions is set to an heuristic estimate of the total cost so that no time is wasted exploring branches that exceed this cost. The possible cluster mappings for the leaf nodes are updated $\hat{\textrm{U}}$ for branch $1$, and these cluster-layers are put onto the queue for updating (Component \nameref{alg:init}).
\begin{algorithm}
\renewcommand\thealgorithm{}
\footnotesize
\begin{algorithmic}[1]
\Procedure{Initialisations}{}
\label{alg:init}
\State $\text{G}[1,0..l_\text{max}]\gets h(\pmb{X}_{l=0..l_\text{max}})$, $\text{G}_\text{min}\gets\text{G}[1,l_\text{max}]$ \label{alg:alg:init1}
\State $\hat{\textrm{U}}\bigl[1,1..|\pmb{X}_\textrm{leaf}|,\pmb{X}_\textrm{leaf}\bigr] \gets 
	f\left(1..|\pmb{X}_\textrm{leaf}|\right)$ \label{alg:alg:init2}
\State Put $\bigl\{1,\hat{\text{U}}[1,1..|\pmb{X}_\text{leaf}|,\pmb{X}_\text{leaf},0,
						\hat{\text{G}}(1,\pmb{X}_\text{leaf}),\text{G}[1,0]\bigr\}$ onto $Q$\EndProcedure
\end{algorithmic}
\vspace*{0cm}
\end{algorithm}

\subsubsection{Pop Queue or Activate Branch}
\label{sss:pop}
Potentially, $Q$ could be treated as a heap to maximise computational efficiency \citep{Lavalle2006}. We use $Pop()$ to refer to removing the lowest cost queue entry, choosing between equicost entries arbitrarily (Component \nameref{alg:pop}). If there are no eligible entries on the queue, i.e.~no branch has $\text{Br}_\text{pl}\geq l$, then a branch is chosen for activation. This makes queue entries on that branch potentially eligible. Activation randomly switches between using $Peek()$ to return, but not remove, the lowest cost entry, or $Peek'()$ to return an entry at the lowest layer on the queue, choosing arbitrarily between entries at the same layer. This facilitates switching between best-first and breadth-first search (Theorem \ref{theorem:leastcost}). After activation, DCMAP tries again to pop the queue if it is not empty; the algorithm only proceeds to the next step (Component \nameref{alg:prune}) if a queue entry was successfully popped.
\begin{algorithm}
\renewcommand\thealgorithm{}
\footnotesize
\begin{algorithmic}[1]
\Procedure{Pop Queue or Activate Branch }{}
\label{alg:pop}
\State $q=\{b\gets b,u(\pmb{X}_{kl}),l,\hat{g}(\pmb{X}_{kl}),\text{G}\} \gets Pop(Q')$
\If{$q=\varnothing$}
		\State $q\gets Peek(Q)$ with probability $\alpha$, otherwise $q\gets Peek'(Q)$ 
		\State $\text{Br}_{pl}(q\$b)\gets|\text{Br}_{pl}(q\$b)|$
\EndIf
\EndProcedure
\end{algorithmic}
\vspace*{0cm}
\end{algorithm}

\subsubsection{Prune Linked Dominated Branches}
\label{sss:prune}
Component \nameref{alg:prune} is essential to achieving computational efficiency in dynamic programming, and effectively removes any entries on the queue relating to that branch. Note that pruning of unlinked branches also happens based on the global cost (Section \ref{sss:layerupd}).
\begin{algorithm}
\footnotesize
\begin{algorithmic}[1]
\Procedure{Prune Linked Dominated Branches }{}
\label{alg:prune}
\State $\pmb{B}\gets j:\text{Br}_\text{lk}[b,l,j,l]=1 \land
							\text{G}[j,l]> 0$
		\State $\forall j\in\pmb{B}: G[j,l] >\min\left({\text{G}[\pmb{B},l]}, \text{G}_\text{min}\right) $, 
				prune $j$ from $\text{Br}_\text{pl}$, $Q$ \label{alg:alg:rmbr}
\EndProcedure
\end{algorithmic}
\vspace*{0cm}
\end{algorithm}

\subsubsection{Propose Cluster Mappings}
\label{sss:proposalscl}
The requirement for contiguous clusters (Definition \ref{defn:kcontig}) means that a node can only be mapped to a new cluster $u_\text{new}$, or the cluster of one or more child nodes $\bf{u}_\text{exist}$. For the former, we can ensure uniqueness and avoid accidentally mapping to an existing cluster by associating a unique label $1..n$ with each node. For the latter, define a priori a $n\times n$ matrix $S:S[i,j]=\{0,1\}$, where $1$ indicates that node $X_i$ can potentially be mapped to the same cluster as node $X_j$. 
To ensure contiguity, Algorithm \ref{alg:sameclust} works backward by identifying, for a given node $X_i$, which nodes $X_j$ at layer $l(X_i)$ or earlier layers can be in the same cluster as $X_i$ that are connected by a common parent node. Note that descendants are child nodes, child of child nodes and so on.
For node $X_j$, $u(X_{i'}) : S[i',j]=1$ are all the node clusters that $X_j$ could be mapped to. To preserve contiguity, node $X_j$ must be such that there are no paths $\pmb{X}_p$ following the directed arcs between $X_i$ and $X_j$ containing any other nodes i.e.~$\pmb{X}_{p,ij}=\{X_i,X_j\}$ or $\pmb{X}_{p,ij}=\varnothing$. The former implies $X_j$ is a parent of $X_i$, and the latter is where both nodes are at the same layer, $l(X_j)=l(X_i)$.
\begin{algorithm}
\footnotesize
\caption{Identifying nodes that can be in the same cluster as a given node $X$.}
\label{alg:sameclust}
\begin{algorithmic}[1]
\Procedure{Samecluster}{}
\State $S[1..n,1..n] \gets 0$
\For{i in 1..n}
	\State $\pmb{X}' \gets \Bigl\{Par(X_i), Descendants\bigl(Par(X_i)\bigr)\Bigr\}$
	\State $S\left[i,j \right] \gets 1, \forall X_j\in\pmb{X}' : \max(|\pmb{X}_{p,ij}|)=2 \lor l(X_j)=l(X_i)$
\EndFor
\EndProcedure
\end{algorithmic}
\end{algorithm}
\begin{lemma}
\label{lemma:sameclustcontig}
When applied iteratively and incrementally from layer $0$ to $l_\text{max}$ (Fig. \ref{fig:overall}), all feasible contiguous cluster mappings (Definition \ref{defn:kcontig}) can be generated using matrix $S$ (Algorithm \ref{alg:sameclust}) to identify all nodes $X_j$ that can be mapped to the same cluster as node $X_i$.
\end{lemma}
\begin{proof} 
The assumption of iterating incrementally over $l=0,1,\ldots,l_\text{max}$ means that at layer $l$, only nodes at layer $l$ or earlier need to be considered for mapping to the same cluster. Additionally, Algorithm \ref{alg:sameclust} only allows nodes $X_i$ and $X_j$ to be mapped to the same cluster if they are directly connected (i.e.~a parent or child of the parent) with no paths going through an intermediary node; this guarantees contiguity, even if $X_i$ and $X_j$ are not in successive layers (Remark \ref{remk:layer}). At each iteration, feasible cluster mappings for node $X_j$ are the clusters of nodes $X_i$ associated with non-zero rows of $S[,j]$, which might be in the same layer or at a child layer.
By Definition \ref{defn:layer} of layers, there must exist at least one parent-child path between each node in one layer and the previous layer, hence, there are no contiguous cluster mappings that cannot be generated by iteratively and incrementally applying $S$ from $l=0$ to $l=l_\text{max}$.
\end{proof}

We demonstrate Algorithm \ref{alg:sameclust} to find $S$ for the simple BN example in Fig. \ref{fig:eg}, comprising three layers and seven nodes:
\begin{equation}
\footnotesize
\begin{bmatrix}
&A 	&B 	&C 	&D 	&E 	&F 	&G\\
 A& 0 & 0 & 0 & 0 & 0 & 0 & 0 \\ 
  B&0 & 0 & 0 & 0 & 0 & 0 & 0 \\ 
  C&0 & 0 & 0 & 0 & 0 & 0 & 0 \\ 
  D&1 & 1 & 0 & 0 & 0 & 0 & 0 \\ 
  E&0 & 0 & 1 & 0 & 0 & 0 & 0 \\ 
  F&1 & 0 & 0 & 0 & 0 & 0 & 1 \\ 
  G&0 & 0 & 0 & 1 & 1 & 0 & 0 \\ 
   \end{bmatrix}
\label{eqn:egsameclust}
\end{equation}
For instance, node $F$ (row $F$) could share the same cluster as nodes $A$, $D$ and itself (Algorithm \ref{alg:sameclust}). Thus, when updating (popping) node $D$ (column $D$) as the search progresses from layer $0$ to $2$, it could be mapped to the current cluster of node $F$ or $G$ or itself. Locally, this could lead to contiguous clusters involving D: $(A,F,D,G)$, $(A,F,D)$, $(F,D)$, $(D,G)$, $(D)$.

The above constrained space of feasible cluster mappings could be pre-computed at initialisation and are enacted in Component \nameref{alg:updateUQ} through $\hat{\textrm{U}}$. Component \nameref{alg:genclust} generates from $\hat{\textrm{U}}$ and particular $k$ and $l$ the cluster-mappings for cost evaluation at each iteration. Intuitively, for a given branch and layer $l$, $\pmb{Z}_1$ represents the nodes that cannot be in any other cluster $k'\neq k$, $\pmb{Z}_3$ are nodes that have already been mapped to another cluster $k'$, and $\pmb{Z}_2$ are the nodes that could be in $k$ or another cluster $k'$. Each of the $N$ unique cluster mapping combinations are stored on separate branches, producing $N-1$ new branches, done by updating data structures and duplicating queue entries from the current branch $b$ for new branches $b'$.
\begin{algorithm}
\footnotesize
\renewcommand\thealgorithm{}
\begin{algorithmic}[1]
\Procedure{Proposing cluster mappings}{}
\label{alg:genclust}
\State $\pmb{Z} \gets\bigl\{X\in\pmb{X}_l:\hat{\textrm{U}}
			\left[b,u(\pmb{X}_{kl}),X\right]= u(\pmb{X}_{kl})\bigr\}$
\State $\left\{\pmb{X}_1,\ldots,\pmb{X}_{N}\right\}\gets 
			\textsc{generatecombos}(\pmb{Z},\text{U},\hat{\text{U}})$

\For{$i$ in $2..N$}
			\State Make new branch $b'$: $\hat{\text{U}}$,$\text{U}$, $\text{G}$, 
				$\text{Br}_\text{pl}(b')\gets -l$ \label{alg:alg:newbranch}
			\State{Duplicate $q'\in Q:q'\$b=b,q'\$l=l$ with $q'\$b=b'$}
				 \label{alg:alg:newbranch2}
		\EndFor

\EndProcedure
\Procedure{generatecombos}{$\pmb{Z},\text{U},\hat{\text{U}}$} 
\State $\pmb{Z}_2\gets\pmb{Z} \setminus \{\pmb{Z}_1,\pmb{Z}_3\} :
	\bigl(\text{U}[b,\pmb{Z}_3]\neq u(\pmb{X}_{kl}), U[b,\pmb{Z}_3]\neq 0\bigr), 
	\bigl(\forall\hat{\text{U}}[b,u\neq u(\pmb{X}_{kl}),\pmb{Z}_1]=0\bigr)$

\State $\pmb{X}_{i=1..N}\gets\Bigl\{\pmb{z}_j\cup\pmb{Z}_1\text{ } \forall\pmb{z}_j\in\text{Combinations}{\pmb{Z}_2\choose 1..|\pmb{Z}_2|} , \pmb{Z}_1\Bigr\}$  \label{alg:alg:combos}
\EndProcedure
\end{algorithmic}
\vspace*{0cm}
\end{algorithm}

Consider a simple example for a given branch and layer with the following possible cluster mappings for each node:
\begin{equation}
\hat{\textrm{U}}\left[b,l(\pmb{X}_i)=l\right] = 
	\begin{bmatrix}
		k	& X_1	& X_2	& X_3 \\
		2 	& 0		& 1		& 1	\\
		3	& 1		& 1		& 1 \\
	\end{bmatrix}
\label{egUhat}
\end{equation}
The feasible combinations are: 
\begin{equation}
u(X_1,X_2,X_3)=\left\{(3,2,2),(3,2,3),(3,3,2),(3,3,3)\right\},
\label{egUhat2}
\end{equation}
noting that $X_1$ can only be mapped to cluster 3; each combination is evaluated on a separate branch.

\subsubsection{Compute Costs}
\label{sss:compute}
Each cluster-mapping proposal is evaluated with respect to the objective function (Component \nameref{alg:compute}) and the cluster mapping proposal is stored via $\text{U}$. 
\begin{algorithm}
\footnotesize
\begin{algorithmic}[1]
\Procedure{Compute Costs }{}
\label{alg:compute}
			\State Compute $\hat{g}(\pmb{X}_i) $ and $\text{G}[b',l] \gets G(\pmb{X}_l)$ 
				(\ref{dpeqng}) and (\ref{dpeqng3}) \label{alg:alg:computeG}
				\State Update $\textrm{U}[b',\pmb{X}_i]=u(\pmb{X}_{kl})$
\EndProcedure
\end{algorithmic}
\vspace*{0cm}
\end{algorithm}

\subsubsection{Layer Updates}
\label{sss:layerupd}
When all the nodes in a single layer are mapped to clusters, it is possible to update $G^*(\pmb{X}_L)$ (\ref{dpeqng}) given the unique cluster mapping on this branch (Lemma \ref{lemma:genclust}). Component \nameref{alg:layerupd} potentially: (i) makes queue entries at layer $l+1$ eligible to update via $\text{Br}_\text{pl}$, (ii) if $l=l_\text{max}$, updates the global minimum objective function cost for global pruning, and (iii) links to other branches with the same cluster mapping for layer $l$ for pruning dominated branches, which occurs at the beginning of each iteration.
\begin{algorithm}
\footnotesize
\begin{algorithmic}[1]
\Procedure{Layer Updates }{}
\label{alg:layerupd}
\If{$\forall X\in\pmb{X}_l:\textrm{U}[b',X]\neq 0$}  \label{alg:alg:layerupd}
						\State Update $\text{Br}_\text{pl}(b')=l+1$ \label{alg:alg:brplupdate}
						\IfThen{$l=l_\text{max}$}{$\text{G}_\text{min}=
						\min{(\text{G}_\text{min},\sum\limits_{l=0}^{l_\text{max}}{\text{G}[b',l]})}$}
						\IfThen{$\forall b'' : \text{Br}_\text{pl}(b'')\geq l+1 \land 
							\text{U}[b'',\pmb{X}_l]=\text{U}[b',\pmb{X}_l]$}{
							$\textsc{link}\left(b',b'',l+1\right)$} \label{alg:alg:brlk}
					\EndIf
\EndProcedure
\end{algorithmic}
\vspace*{0cm}
\end{algorithm}

\subsubsection{Update Possible Cluster Mappings and Queue}
\label{sss:updateUQ}
The final Component \nameref{alg:updateUQ} uses the constrained set of possible cluster mappings (Section \ref{sss:proposalscl}) when updating $\hat{\text{U}}$ for parents of nodes in $\pmb{X}_{kl}$. In addition, it places the possible cluster-layers associated with these parent nodes onto the queue $Q$.
\begin{algorithm}
\footnotesize
\begin{algorithmic}[1]
\Procedure{Update Possible Cluster Mappings and Queue}{}
\label{alg:updateUQ}
\State Generate $u\bigl(X\in Par(\pmb{X}_i)\bigr):u(X)=\{u_\text{new}(X),\pmb{u}_\text{exist}(X)\}$ (Algorithm \ref{alg:sameclust}) 
 						\label{alg:alg:propparind}
					\State $\hat{\text{U}}\bigl[b',u\left(Par(\pmb{X}_i)\right),Par(\pmb{X}_i)
						\bigr]\gets 1$
					\State Put $\bigl\{b',u(Par(\pmb{X}_i),l,
								\hat{g}(\pmb{X}_i),\text{G}[b',l]\bigr\}$ onto $Q$
\EndProcedure
\end{algorithmic}
\vspace*{0cm}
\end{algorithm}

\subsection{Theoretical Analysis}
\label{ss:theoretical}
This section proves analytically that DCMAP is optimal in the sense that it finds all least cost cluster mappings, and does so such that an equally informed algorithm cannot do so in fewer iterations. This is achieved through conditional optimality (derived in Section \ref{sss:condopt}), the exploration of all possible contiguous cluster mappings through branching (Section \ref{sss:branch}), and combined to show the optimality of DCMAP through evaluating non-dominated branches, i.e.~potential cluster mapping solutions, at the earliest opportunity (Section \ref{sss:optimality}). An additional property is derived for super additivity of cost functions for a heuristic to further speed up the search (Appendix \ref{sup:superadd}). 

\subsubsection{Conditional Optimality}
\label{sss:condopt}
Foundational to DCMAP is equation (\ref{dpeqng}) and conditional optimality of a layer cluster mapping, denoted as $u(\pmb{X}_l) = U_l$ for short; the cluster mapping for layers $l'=0,\ldots,l-1$ is denoted as $U_{<l} = U_{l-1}, U_{l-2},\ldots$, and optimal mappings or costs are denoted with a superscript asterisk, e.g.~$G^*$.
\begin{defn}
Given a layer cluster mapping $U_l$, define $U^*_{<l}$ as a conditionally optimal or conditionally non-dominated cluster mapping iff $G^*(\pmb{X}_l|U_l,U^*_{<l}) \leq G(\pmb{X}_l|U_l,U^i_{<l})$ for all possible cluster mappings of earlier layers $U^i_{<l}$. 
\label{defn:condopt}
\end{defn}
Following Definition \ref{defn:dominate}, we say that $U^*_{<l}$ dominates $U^i_{<l}$ if $G^*(\pmb{X}_l|U_l,U^*_{<l}) < G(\pmb{X}_l|U_l,U^i_{<l})$.

\begin{lemma}
\label{lemma:G}
A globally optimal cluster mapping $U^\star=u^\star(\pmb{X})$, where $G^\star(\pmb{X}|U^\star) \leq G(\pmb{X}|U^i)$ for all  other cluster mappings $U^i$, is composed of conditionally optimal $G^*(\pmb{X}_l|U_l,U^*_{<l}) $, layer-based sub-mappings $(U_l,U^*_{<l})$ for each of $l=0,\ldots,l_\text{max}$; i.e.~every layer of $U^*$ is conditionally optimal.
\end{lemma}
\begin{proof}
\label{proof:G}
Equation (\ref{dpeqng}) was derived from dynamic programming to find globally optimal cluster mappings based on optimal mappings between layer $l$ and its latest parent layer $L$. This could be re-written in terms of conditional optimality $G^*(\pmb{X}_{l+1}|U_{l+1},U^*_{<l+1})$ of a mapping at layer $l+1$ which, given a positive, non-zero cost $c(\cdot)$, can only be conditionally optimal if $G^*(\pmb{X}_{l}|U^*_l,U^*_{<l})$ is also conditionally optimal, by definition.
\begin{align}
\footnotesize
&G^*(\pmb{X}_{l+1}|U_{l+1},U^*_{<l+1}) = G^*(\pmb{X}_{l}|U^*_l,U^*_{<l}) + \nonumber\\
&{\sum_{k=1}^{m_{l+1}}{c\Bigl( 
	\pmb{X}_{kl+1},  u(\pmb{X}_{kl+1}), Child(\pmb{X}_{kl+1}), u\bigl(Child(\pmb{X}_{kl+1})\bigr)\Bigr) }}
\label{dpeqng3}
\end{align}

We now prove Lemma \ref{lemma:G} by induction. By Definition \ref{defn:condopt}, the globally optimal objective function cost is equal to the conditionally optimal cost of the last layer $G^\star(\pmb{X}|U^\star) = G^*(\pmb{X}_{l_\text{max}}|U^*_{l_\text{max}},U^*_{<l_\text{max}})$. Note that conditional optimality of $U^*_{<l_\text{max}}$ is defined for a given cluster mapping at the specified layer, hence $U^*_{l_\text{max}}$ refers to the cluster mapping at layer $l_\text{max}$ that minimises $G$ globally.
By equation (\ref{dpeqng3}), a cluster mapping with conditional optimality $G^*(\pmb{X}_{l+1}|U^*_{l+1},U^*_{<l+1})$ at layer $l+1$ requires a conditionally optimal cluster mapping at layer $l$ .
Therefore, by induction, all layers of an optimal cluster mapping must themselves be conditionally optimal.

\end{proof}

\subsubsection{Branching}
\label{sss:branch}
The exploration of all non-dominated cluster mapping solutions, one stored on each branch, is shown through Lemma \ref{lemma:genclust}.
\begin{lemma}
\label{lemma:genclust}
DCMAP can generate all possible contiguous and feasible cluster mapping solutions with each unique mapping stored on a separate branch $b$. A cluster mapping is feasible if every node is mapped to exactly one cluster.
\end{lemma}
\begin{proof}
It follows directly from Component \nameref{alg:genclust} that, given potential cluster mapping proposals for layer $l$ where $\hat{\text{U}}[b,k,]=1$, all feasible and unique cluster mappings are generated and stored on separate branches. Note that, on this branch, any nodes that have already been mapped to a cluster are excluded from the generated mappings via $\pmb{Z}_3$ for feasibility; any nodes that must be mapped to $k$ and have no other mapping proposals are captured via $\pmb{Z}_1$ for feasibility; and combinations of inclusion and exclusion of the remaining nodes are captured via $\pmb{Z}_2$.

From Lemma \ref{lemma:sameclustcontig}, an algorithm that iterates from $l=0$ through to $l=l_\text{max}$ using $S$ from Algorithm \ref{alg:sameclust} will generate all possible contiguous cluster mapping proposals over successive layers, which are stored in $\hat{\text{U}}$ (Component \nameref{alg:updateUQ}). Within each cluster-layer inside each of these layers, Component \nameref{alg:genclust} generates all feasible and unique cluster mappings and stores them onto separate branches at each iteration (i.e.~cluster-layer). Hence, it follows that DCMAP can generate all possible feasible and contiguous cluster mappings, with each unique mapping stored on a separate branch.
\end{proof}

\subsubsection{Optimality}
\label{sss:optimality}
We show that DCMAP will find all the least cost cluster mapping solutions, and that this is done such that an equally informed algorithm cannot do so in fewer iterations in the absence of branch activation (i.e.~$\text{Br}_\text{pl}(b')\gets l$ in Component \nameref{alg:genclust}).
Then, we show the efficiency of branch activation when there are many branches of equal estimated total cost $\hat{g}$.
\begin{theorem}
\label{theorem:leastcost}
DCMAP will incrementally explore branches (i.e.~cluster mapping solutions) from layer $0$ through $l_\text{max}$ to find every least cost $G^\star(\pmb{X},U^\star_i)$ solution.
\end{theorem}
\begin{proof}
Each iteration of DCMAP corresponds to an exploration of a cluster-layer mapping $kl$. When an iteration produces a complete cluster mapping $\text{U}[b',\pmb{X}_l]$ for layer $l$ that is the same as one or more branches $b''$, a link is made between these branches in the same iteration. 
By Definition \ref{defn:condopt} of conditional optimality, any dominated branches sharing the same cluster mapping for that layer $\text{U}[b',\pmb{X}_l]$ can be pruned since by Lemma \ref{lemma:G}, an optimal cluster mapping comprises layer-by-layer conditionally optimal cluster mappings. Pruning of all dominated linked branches happens at the beginning of an iteration (Component \nameref{alg:prune}), so no iterations are wasted exploring a dominated branch. 

In addition to conditional optimality, there is also outright global optimality as captured in $\text{G}_\text{min}$, which affects all branches including those that are not linked. This is updated in the same iteration as when a branch generates a lower cost cluster mapping for the entire BN (Component \nameref{alg:layerupd}). As the objective function cost is non-negative, additive and time-invariant, any branches that exceed $\text{G}_\text{min}$ are globally dominated, even if they have not yet reached $l_\text{max}$, since they cannot possibly lead to a lower cost solution. These are also pruned at the begining of each iteration (Component \nameref{alg:prune}).

By Lemma \ref{lemma:genclust}, DCMAP can generate all possible unique solutions on each branch, but all dominated branches and globally dominated branches are pruned in the process, leaving optimal solutions at algorithm termination, i.e.~when the queue is exhausted. 
It follows that the optimal number of clusters $m$ (\ref{argmineqn}) are also determined.
Branch activation, which only affects the eligibility of $kl$ proposals for popping based on $U_{<l}$ already mapped layers (i.e.~$Q'$), does not affect branch linking and pruning and thus optimality of solutions are preserved. 
\end{proof}
\begin{lemma}
\label{lemma:leastiter}
In the absence of branch activation, let D be an algorithm no more informed than DCMAP that also searches cluster-layers using conditional optimality to find all optimal cluster mappings of a BN. Assume that the objective function and costs are identical: if a cluster-layer was popped by DCMAP, it was also popped by D.
\end{lemma}
\begin{proof}
Suppose the contrary where there is at least one cluster-layer $\kappa\lambda$ on branch $\beta$ popped by DCMAP but not by D. This can only happen if branch $\beta$ was pruned by D as a dominated branch at an earlier iteration, otherwise D cannot be guaranteed to find least cost cluster mappings. If $\beta$ was dominated at an earlier layer $l'<\lambda$, then DCMAP would have found this when completing the mapping of layer $l'$ (Component \nameref{alg:layerupd}), and pruned branch $\beta$ at the earliest opportunity thus never popping $\kappa\lambda$, contradicting the assumption on $\kappa\lambda$ for DCMAP. If $\beta$ is dominated at layer $\lambda$, this implies $\kappa\lambda$ completes layer $\lambda$ and D can only determine this to be dominated by popping $\kappa\lambda$, contradicting the $\kappa\lambda$ assumption for D, hence completing the proof.
\end{proof}
Therefore, it is not possible to find, using an equally informed algorithm, all the least cost cluster mappings in fewer iterations than DCMAP in the absence of branch activation. 
Examples of alternate algorithms for cluster mapping include tree search algorithms like breadth first or depth first search \citep{Lavalle2006}. 
However, the number of iterations needed to find the first optimal solution is also important. 

Consider the impact of pruning on search iterations.
By Definition \ref{defn:condopt} of conditional optimality, pruning of dominated branches can only occur when a layer is completed (Component \nameref{alg:layerupd}). Therefore, changing the order of updating of cluster-layers through branch activation in DCMAP has no effect on the total number of search iterations; it is already optimal (Lemma \ref{lemma:leastiter}). However, branch activation, which evaluates one search branch to completion at a time, can potentially reduce the number of iterations to find the first least cost solution. 
This occurs as cluster-layers in a branch are updated in a semi-greedy ordering since each iteration evaluates cluster-layer proposals with the lowest $\hat{g}(\pmb{X}_i)$ first (Component \nameref{alg:compute}).

Branch activation is especially relevant when there are many queue entries of equal estimated global cost $\hat{g}$, which arises from the combinatorial explosion of possible cluster mappings. 
Consider an order of magnitude approximation of this. Specifically, assume on average $\overline{N}_K$ cluster-layer proposals $k'l'$ at $\overline{N}_P$ parent layers $l'>l$ over $N$ branches (Component \nameref{alg:updateUQ}). 
Note that whenever a new branch $b'$ is made from $b$, queue entries for branch $b$ are copied onto $b'$. 
As a result, the expected number of entries being placed on the queue $\overline{N}_T$ at layer $l$ that share the same $\hat{g}$ as, on average, $\overline{N}_K\overline{N}_P$ other entries, is approximately:
\begin{equation}
\overline{N}_T= \overline{N}\overline{N}_K\overline{N}_P.
\label{eqn:nt}
\end{equation}
For example (\ref{egUhat2}), there are four cluster mappings ($N=4$), $N_K=2$ unique clusters for the first three proposals and one for the last one; assuming $\overline{N}_P=2$, $N_T=14\approx\overline{N}_T=16$.
\begin{remk}
Comparing DCMAP with branch activation with no branch activation, it follows from (\ref{eqn:nt}) that the expected speed-up to update one layer with on average $\overline{N}_K$ clusters is $O(\overline{N}\overline{N}_P)$ i.e.~$O(\frac{1}{\overline{N}\overline{N}_P})^\text{th}$ the number of iterations are needed.
\label{remk:lspeedup}
\end{remk}
Note that this result holds irrespective of the objective function being used.

\section{Case Study}
\label{s:sim}
\begin{figure}[!t]
\centering
\footnotesize
\subfloat[Overall network]{\label{dbn_overall}\includegraphics[width=0.5\columnwidth]{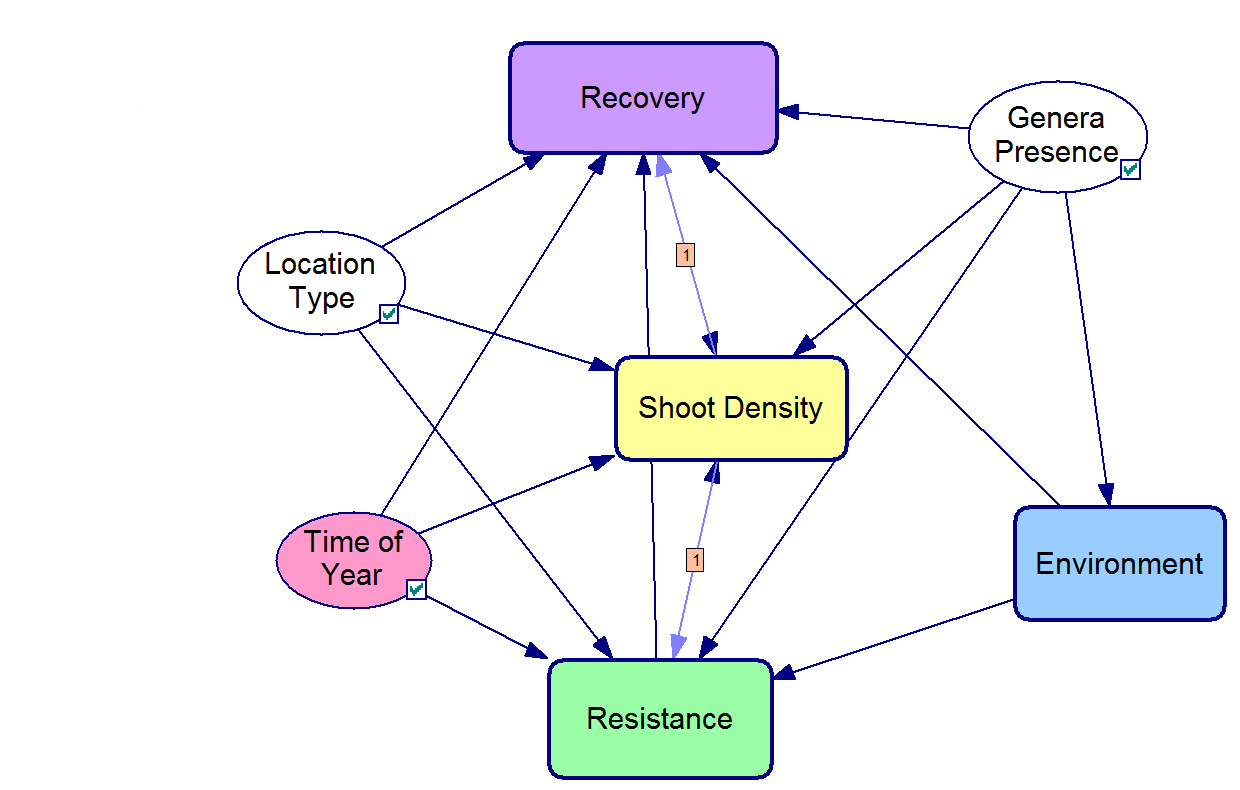}}
\subfloat[Resistance subnetwork]{\label{dbn_resistance}\includegraphics[width=0.5\columnwidth]{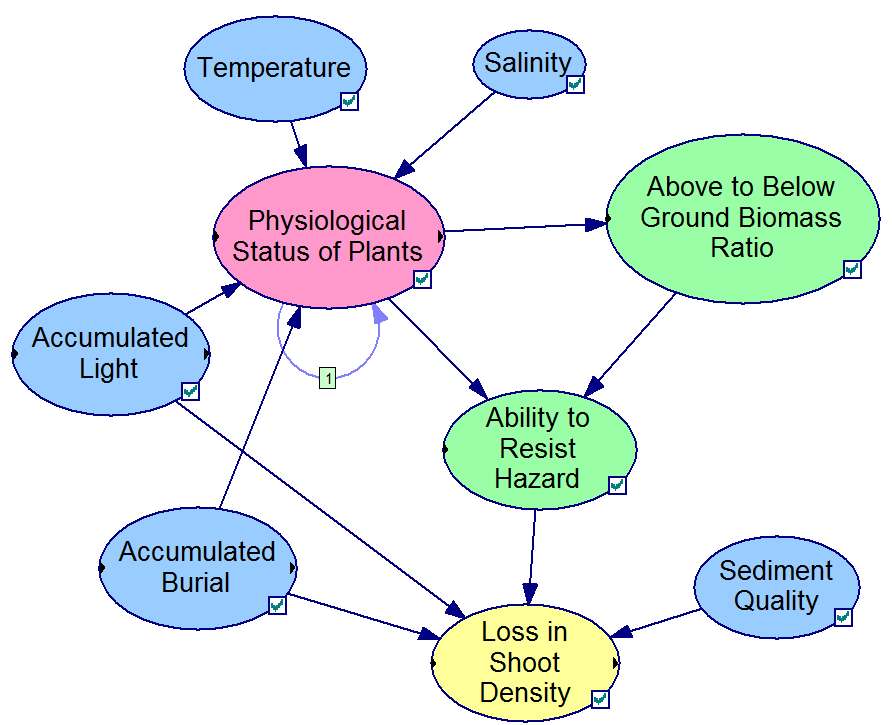}}\\
\subfloat[Recovery subnetwork]{\label{dbn_recovery}\includegraphics[width=0.5\columnwidth]{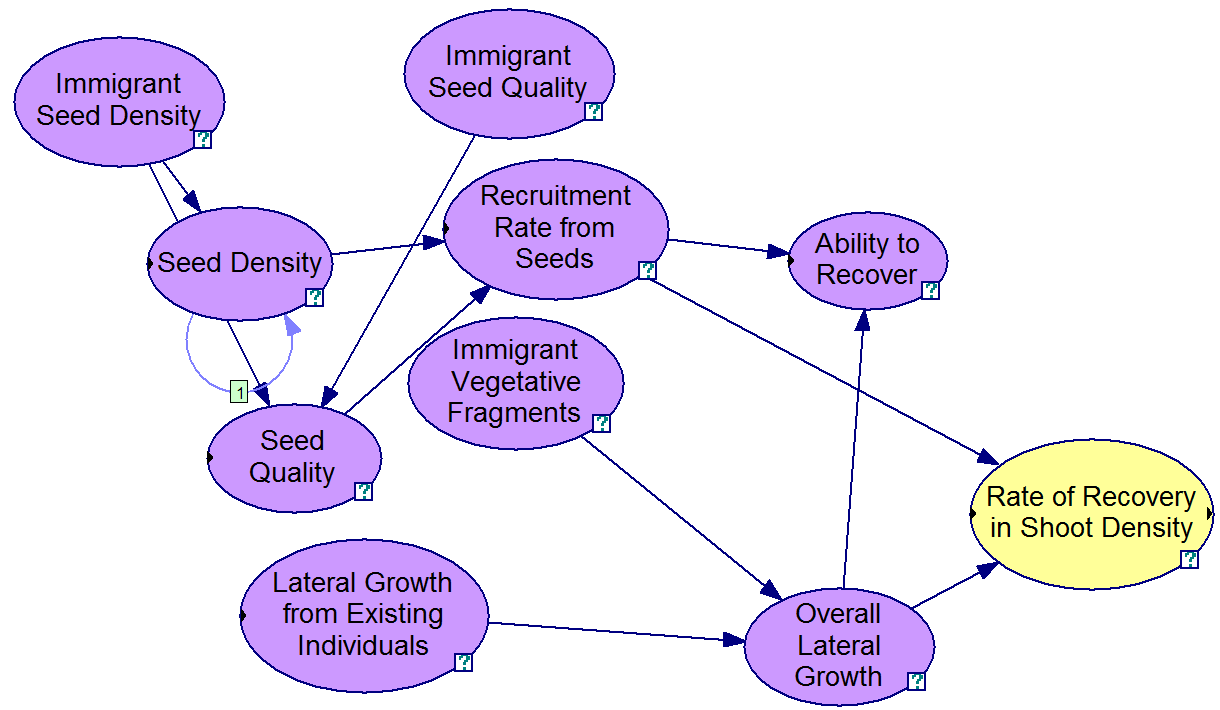}}
\subfloat[Shoot density subnetwork]{\label{dbn_shoot}\includegraphics[width=0.5\columnwidth]{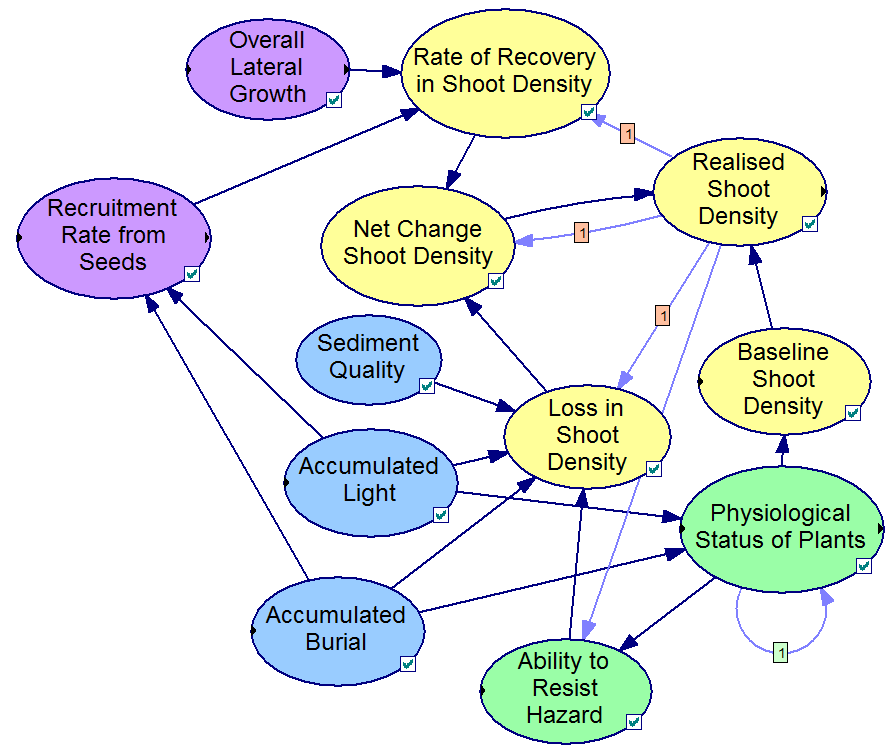}}
\caption{Overall seagrass DBN model \citep{Wu2018} (Fig. \ref{dbn_overall}). Nodes are ovals and arrows denote conditional dependence between a parent and child node in the same time-slice. Where an arrow is labelled with a $1$, the child node is in the next time-slice. Rounded rectangles denote subnetworks. Yellow nodes relate to loss and recovery in shoot density, purple nodes to recovery, green nodes to resistance, blue nodes to environmental factors and pink for all other nodes. The environment subnetwork is not shown as it is a disjoint subnetwork of the blue nodes, which are effectively inputs to the model.} 
\label{seagrassdbn}
\end{figure}

We use as a case study the complex systems DBN model for seagrass resilience, described briefly in Section \ref{s:intro}. This model, which has 25 discrete nodes per time-slice (Fig. \ref{seagrassdbn}) and 96 time slices \citep{Wu2018}, captures biological, ecological and environmental dynamics and their interactions to predict the impact on resilience of a series of dredging stressors and its cumulative effects over time. It demonstrates small world or power law connectivity (Table 3 and 4) with a few nodes having comparatively many parent or child nodes, and the majority having few \citep{Watts1998}; most nodes had two to four states, with a few having ten or more states. \cite{Wu2018} incorporated expert knowledge and multiple data sources (e.g.~experimental and monitoring data) to inform the model; however, the computation time prohibited the use of Bayesian inference, which enables better data integration and estimation of uncertainty. 
\begin{table}
\footnotesize
\caption{Distribution of in- and out-degree (degree and number of nodes with that degree) in the one time-slice DBN.}
\label{tab:inoutdeg}
\centering
\begin{tabular}{|cccccccc|}
\hline\hline
In-degree	& 0 & 1 & 2 & 4 & 5 & 6 & 7\\
\#nodes	& 9 & 2 & 8 & 3 & 1 & 1 & 1\\
\hline
Out-degree& 0 & 1 & 2 & 3 & 4 & 6 & 9\\
\#nodes	& 2 & 14 & 3 & 3 & 1 & 1 & 1\\
\hline\hline
\end{tabular}
\end{table}

For the two time-slice version of this DBN with 50 nodes, the in- and out-degrees are even higher than the one time-slice DBN (Table 4).
\begin{table}
\footnotesize
\caption{Distribution of in- and out-degree (degree and number of nodes with that degree) two time-slice DBN.}
\centering
\begin{tabular}{|cccccccccc|}
\hline\hline
In-degree	& 0 & 1 & 2 & 3& 4 & 5 & 6 & 7 & 8\\
\#nodes		& 18 & 4 & 10 & 6 & 5 & 2 & 2 &2 & 1\\
\hline
Out-degree	& 0 & 1 & 2 & 3 & 4 & 6 & 7 & 8 & 9\\
\#nodes		& 3 & 27 &7 & 6 & 2 & 1 & 1 & 1 &2\\
\hline\hline
\end{tabular}
\label{tab:inoutdeg2}
\end{table}

\subsection{Computation Cost as an Objective Function}
\label{ss:simcomp}
The objective function is defined in terms of the computational cost associated with BN inference (\ref{sumprod1}) as illustrated in Section \ref{ss:eg}. Recall that the objective function is defined recursively with respect to a cost function $c$ (\ref{dpeqng}), hence application of DCMAP to different use cases involves the definition of an appropriate cost function $c$. 

Without loss of generality, consider a discrete BN. We formulate a backwards probability measure $P^b_{kl}$ for computing inference akin to the rearrangement of Equation (\ref{sumprod1}), which forms the basis of inference techniques such as variable elimination \citep{Daly2011}.  
\begin{equation}
P^b_{kl}=\prod_{\forall X_i\in X_{kl}} \sum_{X_i} P\bigl(X_i|Par(X_i)\bigr) \prod_{X_{kl'}\in Child(X_{kl})} \sum_{\forall X \notin X_{kl'}} P^b_{kl'}
\label{pb}
\end{equation}
$P^b_{kl}$ comprises propagation from child cluster-layer $P^b_{kl'}$, marginalising (i.e.~summing) out any nodes that are not in $\pmb{X}_{kl}$, multiplication by the conditional probabilities of the nodes in $\pmb{X}_{kl}$, and finally marginalisation in preparation for propagation to the next antecedent cluster-layer. It is a recursive relation and for $l'=0$, $P^b_{kl'}=1$ since a cluster-layer comprising leaf nodes does not have child nodes, by Definition \ref{defn:layer}. In the context of $c\Bigl(\pmb{X}_{kl'},  u(\pmb{X}_{kl'}), Child(\pmb{X}_{kl'}), u\bigl(Child(\pmb{X}_{kl'})\bigr)\Bigr)$ (\ref{dpeqng}), the limits of the products and sums incorporates the effect of cluster mappings $u$, and $P(X_i|Par(X_i))$ and $P^b_{kl'}$ are defined with respect to cluster-layer and child cluster-layer nodes. 

Define $\mathcal{F}()$ as a normalised count of the number of clock cycles associated with performing all the multiplications and additions, where each multiplication is assumed to have a normalised count of $1$ cycle, each addition $0.6$ and each division (i.e.~ratio) $3$, based on Intel Skylake CPU latencies \citep{Fog2022}. Hence,
\begin{equation}
c\Bigl(\pmb{X}_{kl'},  u(\pmb{X}_{kl'}), Child(\pmb{X}_{kl'}), u\bigl(Child(\pmb{X}_{kl'})\bigr)\Bigr) = \mathcal{F}\bigl( P^b_{kl} \bigr)
\label{compcosteqn}
\end{equation}

\subsubsection{Results and Discussion}
\begin{figure}
\centering
\includegraphics[width=\columnwidth]{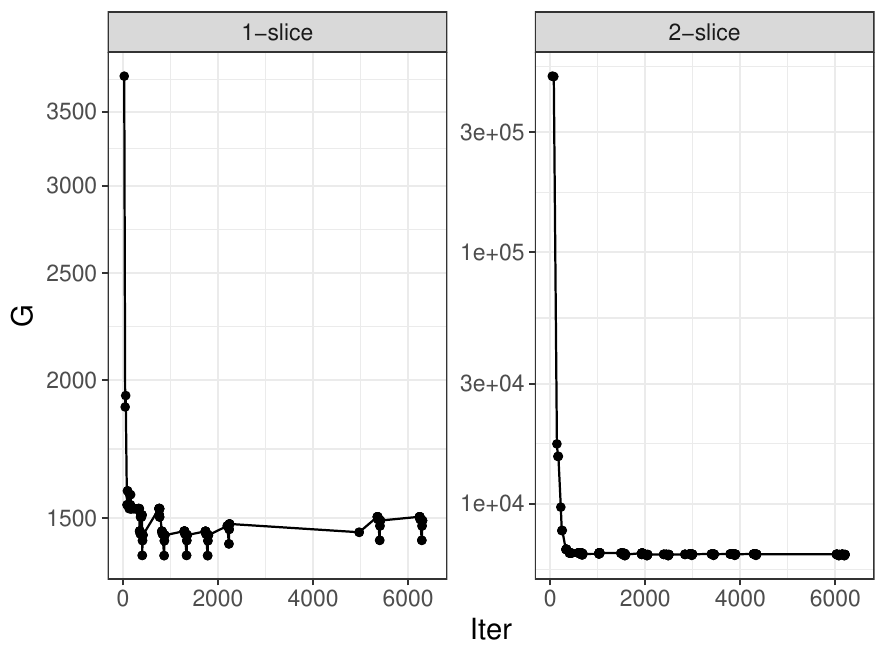}
\caption{Cost $G$ of solutions found and the iteration of their discovery for both the one and two time-slice DBNs.}
\label{fig:iters}
\end{figure}

For one time-slice of this DBN, DCMAP finds a total of 4 optimal cluster mapping solutions, with an optimal cost of 1386.6 compared to a naive heuristic cost of 34444; the first optimal solution was found at iteration 865 and the last one at iteration 1782 (Fig. \ref{fig:iters}). Over 20 runs, the mean iteration for the first optimal solution was 856 $(\text{95\% CI } 852,866)$. In contrast, the two time-slice DBN had 10 least cost solutions, the first found at iteration 1569 and last at iteration 6190 (Fig. \ref{fig:iters}), with an optimal cost of 6220.8 compared to the naive heuristic cost of 3148386. Over 20 runs, the mean iteration for the first optimal solution was 1568 $(1566,1581)$. 
An $\alpha$ probability value of $0.6$ was selected to achieve a balance of best first and breadth first search with a slight preference to the former.
These empirical results suggest a hypothesis that the complexity of finding the first optimal solution is $O(n^2)$ where $n$ is the number of nodes, and that of finding all the least cost solutions is $O(an^2)$ for positive integer $a$; this particular case study suggests a value of $a=3$.
Additionally, DCMAP rapidly converged onto near-optimal solutions (Fig. \ref{fig:iters}), and regularly returned cluster mapping solutions (Fig. \ref{fig:iters}), on average returning a solution every 57 and 24 iterations for the one and two time-slice versions of the study.

\begin{figure}[!h]
\centering
\includegraphics[width=0.9\columnwidth]{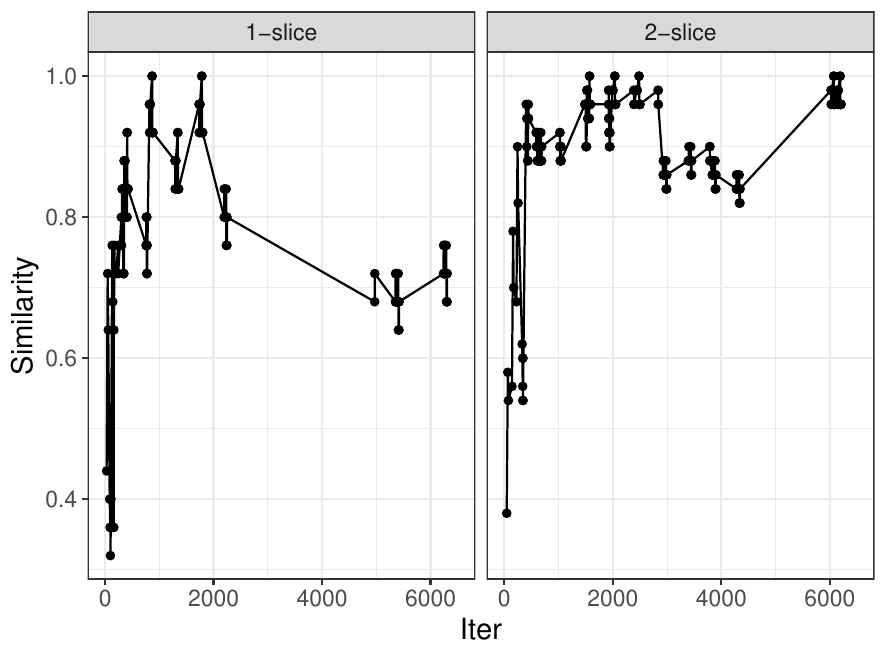}
\caption{Similarity of solutions to the set of optimal solutions and their iteration of discovery for both the one and two time-slice DBNs.}
\label{fig:similarity}
\end{figure}

Furthermore, we could compare the similarity of cluster mappings to optimal cluster mappings using nodes assigned to the same cluster rather than nodes assigned to the same cluster label, as labels could vary depending on its child clusters when assigning a node to an existing cluster (Section \ref{sss:proposalscl}). This could be done by creating a $n\times n$ matrix where $Y[i,j]\in\{0,1\}$ for each cluster mapping, where a 1 indicates that those two nodes are in the same cluster. This way, the similarity of a solution $Y$ compared to an optimal solution $Y^\ast$ could be computed as the dot product $Y \boldsymbol{\cdot} Y^\ast$, divided by the number of ones in $Y^\ast$. Here, we use the maximum similarity between $Y$ and the set of optimal solutions $\pmb{Y}^\ast$. For the one and two time-slice and DBN, there was both rapid convergence in terms of cost (Fig. \ref{fig:iters}) and approximately 80\% or more similarity (Fig. \ref{fig:similarity}) to optimal solutions in just a few hundred iterations.  

In addition to the empirical results above and theoretical computational efficiency of DCMAP (Lemma \ref{lemma:leastiter}), the performance of DCMAP can be further contextualised through comparison with breadth first search using layers and clusters (Section \ref{sss:layers} and \ref{sss:clusters}). On a simple graph such as the illustrative example (Fig. \ref{fig:eg}), DCMAP and breadth first search require similar computation, finding all three optimal solutions at iterations 60, 99 and 100, versus iterations 54, 84 and 86, respectively. However, breadth first search does not scale well with graph complexity (i.e.~size and layers) and is unable to find even one least cost solution for the one time-slice case study after 9600 iterations and the lowest cost solution was 1518.4; DCMAP found an optimal solution of cost 1386.6 at iteration 865 and had found all four optimal solutions by iteration 1782.  

The size of the search space computed using (\ref{ssseqn}) was $1.51\times10^{21}$ and $9.91\times10^9$ for the two and one time-slice DBNs, respectively and, as might be expected given Lemma \ref{lemma:genclust}, all branch solutions are unique (i.e.~no two branches with the same solution). Such exponential growth in search space size likely contributes to the inability for breadth first search to scale with more layers and nodes. In addition, such a large space of cluster mapping solutions contributes to a greater potential for benefit from optimisations, where the optimal solutions had 4\% and 0.2\% the cost of naive solutions, which corresponds to a computational speed-up of 22.8 and 426 in the one and two time-slice case studies, respectively. Therefore, DCMAP facilitates substantially more computationally efficient inference by using clustering in our complex systems seagrass DBN case study.

\section{Discussion}
\label{s:discussion}
%
Clustering has been applied to improve the computational feasibility of BN learning via constraints, scoring functions (i.e.~using a cost function) or both \citep{Huang2022}, and the computational feasibilitiy of inference through heuristics and local optimisation or both \citep{Forouzan2015}. 
However, algorithms that explicitly focus on forming optimal clusters in the context of BN inference are missing. This incurs additional complexity in the form of cost dependency, where the local cost of a cluster is dependent on the mapping of connected nodes and the global cost is typically the sum of local costs. 
The proposed, general optimisation of BN clusters, i.e.~with arbitrary criteria, advances the field by: (i) providing a benchmark for validating approximate methods against the globally optimal solution, (ii) enabling future research and development of cluster-based methods for inference including anytime optimisation \citep{Mateescu2010} and criteria such as entropy \citep{Lin2020} (Section \ref{ss:entropy}), and (iii) finding clusters of homogeneous regions in a system with changing dynamics \citep{Wu2018}.

Previous BN literature has used decomposability of scoring functions, such as through additive costs, to allow for local search procedures and/or optimisation of the score function; some even use dynamic programming and graph layers \citep{Friedman1999, Kojima2010, Lu2021}.
Our work builds upon the application of dynamic programming and layers in BNs to facilitate global optimisation with arbitrary non-negative cost functions and cost dependency. This latter property has not been explicitly studied in the existing BN literature.

The proposed algorithm, DCMAP, is proven to find all of the least cost cluster mapping solutions through conditional optimality of layers and branching (Section \ref{sss:optimality}). In addition, it is theoretically shown that an equally informed algorithm cannot find these solutions in fewer iterations. 
This is important as optimal clustering is NP-hard in general \citep{Buluc2016} and cost dependency would exacerbate this complexity. For instance, a breadth first search on the same one time-slice case study is unable to find any optimal solution even with ten times more iterations than DCMAP. This arises even though the number of iterations needed is similar to DCMAP for the simple illustrative BN (Fig. \ref{fig:eg}). 

\subsection{Case Study Discussion}
We developed a case study of BN clustering where clusters were used as a means to compute (i.e.~infer) the marginal probabilities of a complex systems seagrass DBN. By explicitly defining the cost function as the computational cost of inference given a cluster mapping, a speed-up of 22.8 and 426 times was achieved for a one and two time-slice version of the DBN, respectively.  
In itself, a substantial reduction of computational complexity through clustering is useful as BN inference plays a key role in prediction and model learning, and methods for that such as Markov Chain Monte Carlo (MCMC) \citep{Gelman2013} can require thousands of iterations. MCMC can also help better quantify uncertainty for parameter and structure learning of BNs \citep{Daly2011} and DBNs \citep{Shafiee2020} through the Bayesian framework. 
Explicit optimisation of BN clusters using DCMAP could be a key enabler for future research in BN learning and updating, especially for complex systems DBNs like our seagrass case study.

Although our case study was for a discrete BN with between two and fourteen states per node with corresponding CPTs, the proposed DCMAP algorithm equally applies to discrete or continuous or hybrid BNs, DBNs, Object Oriented Bayesian Networks (OOBNs) \citep{Koller2013} and Markov models in general, all of which can be represented as a DAG; the key adaptation required is the development of an application-specific cost function. 
Note that dynamic and templated models alike can be unrolled \citep{Wu2018} into a BN and cluster mapping performed on the result. In practice, for a $k^{th}$ order dynamic model, running cluster mapping on $k$ slices is sufficient to find optimal clusters that can then be replicated over all slices. 

We have theoretically shown that an equally informed algorithm cannot find all optimal solutions in fewer iterations, further study is needed to better understand the computational complexity of DCMAP to find the first optimal solution and how it might vary with different BN graph structures and cost functions. 
Early results from the case study suggest a complexity of $O(n^2)$ iterations to find the first least cost cluster mapping and $O(an^2)$ iterations for all optimal solutions, where $n$ is the number of nodes and $a$ some positive integer. This aligns with previous findings of $O(n^2)$ complexity for clustering in the absence of cost dependency \citep{Zheng2020}. 
However, the computational complexity could potentially vary significantly depending on the cost function employed and also the structure of the network (e.g.~dense versus sparse). 
Additionally, it could also vary with the $\alpha$ probability for switching between best first and breadth first search (Section \ref{sss:pop}), especially for near-optimal solutions. 
These are all avenues for future research.

Despite the theoretical optimality of DCMAP, scalability in practice is a challenge for large networks with thousands of nodes. Potentially, the template behind a DBN or OOBN like the two-slice DBN case study, and/or near-optimality could be investigated as approaches for large networks. Better understanding of the convergence characteristics and early termination of clustering, including in the impact of cost functions and network structures, is needed. Additionally, even though proposed cluster mappings are always unique (Lemma \ref{lemma:genclust}), label switching is an area for future study to better understand the extent of the issue and strategies for mitigation if needed.

\subsection{Future Work: Entropy}
\label{ss:entropy}
In light of the generalisability of DCMAP given its ability to optimise for arbitrary, non-negative and dependent cost functions, one key area of future work is in the development of objective functions for specific usage scenarios. One such possibility is entropy, which for BNs is defined generally as the sum of the entropy associated with each scenario over all possible scenarios, where a scenario might be defined in terms of one or more nodes \citep{Scutari2024,Parhizkar2018}. Consider a discrete BN. The entropy associated with evidence for a given scenario indexed by $k$ is:
\begin{equation}
H_k = -\sum_{k=1}^{n^k} p_k \sum_{i=1}^n \sum_{j=1}^{n^s_i} p_{ijk} log(p_{ijk})
\label{entropy}
\end{equation}
where we represent scenarios as clusters $k$, $n^k$ is the total number of clusters, $p_k$ is the marginal probability associated with that scenario, $i$ indexes nodes and $n$ is the total number of nodes, $j$ indexes states where $n^s_i$ is the number of states for node $i$, and $p_{ijk}$ is the marginal probability of node $i$, state $j$ and scenario $k$.

Consider a special case of clustering where specific evidence scenarios are to be applied at one or more nodes and at one or more points in time in a DBN, with the goal of minimising global entropy. Thus, each node can be assigned to one of two clusters: (cluster 1) observed using virtual evidence $\delta(X')$ \citep{Pearl1988} where each observation is independent, or (cluster 2) not observed. Note that virtual evidence provides greater flexibility in modelling uncertainty compared to hard evidence where a state is observed or not. Dependent cluster mapping is necessary to optimise over all possible combinations of observed nodes and their interdependencies, which arise from propagation of evidence for inferring marginal probabilities for each scenario $p_{ijk}$ (\ref{entropy}). DCMAP can approximate this marginal probability $p_{ijk}$ for every node $X_i$ via backwards propagation with evidence to layer $l$ and no evidence from forward propagation:
\begin{equation}
p_{ijk} = \prod_{\forall X' : l(X')\leq l} \sum_{X' \notin X_i} P\bigl(X'|Par(X')\bigr)\delta(X') P^f_{l+1}
\label{pbentropy}
\end{equation}
where forwards propagation $P^f_{l+1} = \prod_{\forall X' : l(X')> l} \sum_{X'} P\bigl(X'|Par(X')\bigr)$, which assumes no evidence since these nodes have not been mapped yet, and could be pre-computed. Due to independence of evidence, cluster proposal (Section \ref{sss:proposalscl}) only involves assigning one node at a time $\pmb{X}_{kl}=X_i$ to being observed or not, and $p_k$ is equal to $p_{ijk}$ in equation (\ref{pbentropy}) in the absence of evidence for $X_i$. These enable the computation of an additive objective function $G$ which sums entropy over all evidence scenarios $k$ via $H_k$ (\ref{entropy}). 

Clusters have been used to improve the computational efficiency of inference, and learn conditional probabilities and BN structures \citep{Daly2011, Lu2021}; potentially, optimal clusters can further enhance these tasks. In addition, using cost functions such as entropy, clusters could also have a practical interpretation. In non-homogeneous Hidden Markov Models or DBNs for instance, clusters could be interpreted as homogeneous portions of a non-homogeneous system, akin to changepoint detection or model switching \citep{Shafiee2020, Boys2000}. Potentially, optimisation of clusters with respect to entropy could contribute to new methods in BN learning and prediction, and is a promising area for future research that is enabled by DCMAP.
\printbibliography

\appendix
\section{Appendix: Simple Example Comparison of Clustering and Computation Cost}
\label{app:eg}
To illustrate how clustering can support DAG inference, consider an example BN (Fig. \ref{dcmap_clusts}) and the fundamental task of finding posterior distributions, specifically $P(X,\pmb{E}), \forall X\in\pmb{X}$, Equation (\ref{sumprod1}).
\begin{figure}
\centering
\footnotesize
\subfloat[DCMAP Clusters]{\label{dcmap_clusts}
	\begin{tikzpicture}[>={Latex[width=2mm,length=3mm]},
						every node/.style={circle,thick,draw},
						every edge/.style={draw=black,thick}]
		\node (F) [] at (0,0) {F};
		\node (G) [fill=black,text=white, below =of F]  {G};
		\node (D) [fill=black,text=white, left =of G] {D};
		\node (E) [fill=gray!85, below =of D] {E};
		\node (B) [fill=gray!25, left =of D] {B};
		\node (A) [ above =of B] {A};
		\node (C) [fill=gray!60, below =of B] {C};
	
		\path [->] (D) edge (G);
		\path [->] (E) edge (G);
		\path [->] (A) edge (F);
		\path [->] (A) edge (D);
		\path [->] (B) edge (D);
		\path [->] (C) edge (E);
	
		\node[plate=Layer 0, inner sep=10pt, fit=(F) (G) ] (plate0) {};
		\node[plate=Layer 1, inner sep=10pt, fit=(D) (E) ] (plate1) {};
		\node[plate=Layer 2, inner sep=10pt, fit=(A) (B) (C)] (plate2) {};
	
	\end{tikzpicture}
}\hspace{0.8cm}
\subfloat[Jointree Clusters]{\label{jointree_clusts}
	\begin{tikzpicture}[>={Latex[width=2mm,length=6mm]},
						every node/.style={circle,thick,draw},
						every edge/.style={draw=black,thick},
						square/.style={regular polygon,regular polygon sides=4}]
		\node (AF) at (0,0) {AF};
		\node (ABD) [right= 1.6cm of AF] {ABD};
		\node (DEG) [below= 1.5cm of ABD] {DEG};
		\node (CE) [left= 1.6cm of DEG] {CE};

		\draw (AF) -- (ABD) node [midway, fill=white, square] {A};
		\draw (ABD) -- (DEG) node [midway, fill=white, square] {D};
		\draw (CE) -- (DEG) node [midway, fill=white, square] {E};
	\end{tikzpicture}
}\hspace{1cm}
\subfloat[Buckets]{\label{bucket_clusts}
	\begin{tikzpicture}[>={Latex[width=2mm,length=6mm]},
						every node/.style={circle,thick,draw},
						every edge/.style={draw=black,thick},
						square/.style={regular polygon,regular polygon sides=4}]
		\node (G) at (0,0) {G};
		\node (E) [below= 0.25cm of G] {E};
		\node (D) [below= 0.25cm of E] {D};
		\node (C) [below= 0.25cm of D] {C};
		\node (B) [below= 0.25cm of C] {B};
		\node (A) [below= 0.25cm of B] {A};
		\node (F) [below= 0.25cm of A] {F};

		\draw (G) edge [bend left] (E);
		\draw (G) edge [bend left] (D);
		\draw (E) edge [bend right] (D);
		\draw (E) edge [bend right] (C);
		\draw (D) edge [bend left] (B);
		\draw (D) edge [bend left] (A);
		\draw (B) edge [bend right] (A);
		\draw (A) edge [bend right] (F);
	\end{tikzpicture}
}
\caption{Example network showing nodes, directed edges, and layers (Fig. \ref{dcmap_clusts}). Shading denotes clusters (Fig. \ref{dcmap_clusts}) obtained using DCMAP, whereas the jointree \citep{Lauritzen1988} comprises clusters (circles) and sepsets (squares) (Fig. \ref{jointree_clusts}), and Bucket Elimination \citep{Dechter1999} uses buckets (Fig. \ref{bucket_clusts}). Assume all nodes have binary states.}
\label{fig:eg_a}
\end{figure}
%
%
Assuming binary states for all nodes, the Conditional Probability Table (CPT) for each node with no parents (nodes $A$, $B$, $C$) is a $2\times 1$ matrix, e.g.~$P(A)=[p_a,p_{\bar{a}}]^T$ for states $A=a$ and $A=\bar{a}$. For $G$ and $F$ the CPT is a $2\times 2$ matrix, e.g.~$P(F|A)=\begin{bmatrix} p_{af} &p_{\bar{a}f}\\ p_{a\bar{f}} & p_{\bar{a}\bar{f}}\end{bmatrix}$ where $p_{af}$ is the probability of state $F=f$ given state $A=a$. $P(D|A,B)$ is a $2\times 4$ matrix. Multiplication of CPTs (\ref{sumprod}) containing different nodes, i.e.~variables, grows the dimensionality of the resultant matrix with associated exponential growth in computation; in the discrete case, the resultant matrix contains all possible combinations of node states. We ignore evidence terms for notational simplicity, noting that evidence can be easily incorporated using $\delta(X_j)$ with node $X_j$ (\ref{sumprod}). Using (\ref{sumprod1}), the posterior probability $P(F)$ is:
\begin{align}
\footnotesize
P(F) &= \sum\limits_{A,B,C,D,E,G}{P(A)P(B)P(C)P(D|A,B)P(E|C)P(F|A)P(G|D,E)}\label{egjoint_a}\\
P(F) &= \sum\limits_A{P(F|A)P(A)} \sum\limits_B{ P(B)} \sum\limits_C{P(C)} \sum\limits_D{ P(D|A,B)}\sum\limits_E{P(E|C)}\sum\limits_G{ P(G|D,E)}\label{egbucket_appendix}
\end{align}
The computation of (\ref{egjoint_a}) produces a matrix of $7$ dimensions, one for each node $A$ through $F$, with corresponding exponential growth in memory ($2^7$ joint probability values) and computation associated with multiplication and marginalisation, aka product and sum.
In contrast, the maximum dimensionality of (\ref{egbucket_appendix}) obtained through Bucket Elimination (BE) \citep{Dechter1999} is $4$, occurring at bucket $D$ (Fig. \ref{bucket_clusts}). Intuitively, (\ref{egjoint_a}) can be rearranged, because of commutativity, to minimise computation by marginalising (summing) out nodes as early as possible (\ref{egbucket_appendix}). BE in effect employs dynamic programming to optimally compute the posterior marginal probability of a given target node, specified as part of a node ordering. In Fig. \ref{bucket_clusts}, the specified order is (top to bottom) $G$, $E$, $D$, $C$, $B$, $A$, $F$. 

Assume that the addition of two numbers has a computational cost of $0.6$ compared to $1$ for multiplication and $3$ for divison (i.e.~ratio), based on floating point operation latencies for the Intel Skylake CPU as an example \citep{Fog2022}). Consider the computation of $\lambda_G(D,E)=\sum_G{ P(G|D,E)}$ in bucket $G$ of Fig. \ref{bucket_clusts}, which is the last term in (\ref{egbucket_appendix}). Nominally, this comprises multiplication of $1$ by $P(G|D,E)$ which, given every possible combination of binary states, requires $2^3=8$ multiplication operations and $4$ addition operations (one per combination of $D$ and $E$ states). Thus, the output $\lambda_G(D,E)$ is computed at a cost of $8+6\times 0.4=10.4$, and only has two dimensions (i.e. $D$ and $E$) and a joint probability distribution size of $2^2$. Bucket $E$ computes $\lambda_E(C,D)=\sum_E{P(E|C)}\lambda_G(D,E)$, requiring $2^2\times 2$ multiplication operations since there is an overlap in the $E$ dimension between $P(E|C)$ and $\lambda_G(D,E)$. Marginalisation again needs $4$ addition operations. The total computational cost is 64.4 to find $P(F)$ (\ref{egbucket_appendix}); finding the posterior for all seven nodes would incur a cost around 448, depending on input node ordering for BE.

As might be expected, other exact inference methods including Pearl's belief propagation and polytrees, and \cite{Lauritzen1988}'s widely-used clustering algorithm share similarities with BE as all evaluate (\ref{sumprod1}) exactly \citep{Dechter1999}. However, clustering \citep{Lauritzen1988} has a distinct advantage as it does not require a pre-specified node ordering and has less computational cost compared to BE when finding the posterior $P(X,\pmb{E})$ for every node. This is achieved by performing inference in two steps: (i) forward-backward propagation along the jointree (Fig. \ref{jointree_clusts}), and (ii) marginalisation of clusters to get $P(X,\pmb{E})$; note that the posterior distribution given evidence $P(X|\pmb{E})$ is obtained by normalising $P(X,\pmb{E})$. The jointree, comprising clusters connected by sepsets, is derived from the DAG (Fig. \ref{dcmap_clusts}) using a heuristic process known as triangulation. forward-backward propagation involves: (a) projection from one cluster to a connected cluster via the sepset, and (b) absorption of the ratio of the new to old sepset into the receiving cluster. For example, projection of cluster $AF$ to $ABD$ (Fig. \ref{jointree_clusts}) involves computing $\phi_{A} = \sum_A{\phi_{AF}}$, $\phi_{AF}=P(F|A)P(A)$ comprising one nominal multiplication by $P(A)$ then another by $P(F|A)$, then marginalisation for a cost of $2+4+2\times 0.6 = 7.2$. Absorption into $ABD$ updates $\phi_{ABD}=\phi_{ABD}\frac{\phi_A}{\phi^\textrm{old}_A}$, $\phi_{ABD}=P(D|A,B)$ incurring a ratio and multiplication cost of $2\times 3 + 8=14$. After forward-backward propagation, the posterior for $A$ is obtained from cluster $AF$ by $\sum_F{\phi_{AF}}$ for a cost of $2\times 0.6=1.2$; overall the cost to infer all nodes is just 132.8, which is substantially less than BE. 

The above examples of BE and clustering \citep{Lauritzen1988} show how cluster-like structures (e.g. buckets, clusters, sepsets) can facilitate more computationally efficient inference in a DAG; note that these structures differ from our definition of a cluster (Section \ref{s:intro}). Computational efficiency is critical for MCMC-based inference \citep{Gelman2013}, as there is a need to run potentially thousands or hundreds of thousands of iterations of the models. Importantly, MCMC can help better quantify uncertainty for parameter and structure learning of BNs \citep{Daly2011} and DBNs \citep{Shafiee2020}. 
Potentially, explicit optimisation of clusters using criteria associated with inference, represented as a cost, could further improve computational efficiency, make tractable MCMC for complex models, or facilitate new methods for inference such as finding homogeneous portions of a non-homogeneous system. However, this is a non-trivial task as the local cost of a cluster is dependent on the cluster mapping of connected nodes. 

\section{Appendix: Super-additive Objective Functions}
\label{sup:superadd}
A further consideration is that heuristics could also be applied to filter the proposal of cluster mappings $(k,l)$ in addition to the prioritisation of cluster-layers for popping via $\hat{g}$. The former are typically contingent on the specific objective function being optimised. For instance, consider optimisation of DBN computation cost, where including an extra node in a cluster requires multiplication and marginalisation operations (Section \ref{ss:eg}) that exceed the cost of doing so separately, so super-additivity might apply:
\begin{defn}
Using(\ref{dpeqng}), we say that a cost function is super-additive if for any two nodes $\{X_1,X_2\}$ in the same layer $l$:
\begin{equation}
c\left(\{X_1,X_2\},\ldots\right) > c\left(X_1,\ldots\right) + c\left(X_2,\ldots\right).
\label{eqn:superadd}
\end{equation}
\label{defn:superadd}
\end{defn}
\vspace{-0.85cm}
It follows from Definition \ref{defn:superadd} that:
\begin{align}
&c\left(\{X_1,X_2\},\ldots\right) + c\left(X_3,\ldots\right)
> \nonumber\\ &c\left(X_1,\ldots\right) + c\left(X_2,\ldots\right) + c\left(X_3,\ldots\right)\nonumber\\
&c\left(\{X_1,X_2,X_3\},\ldots\right) > c\left(\{X_1,X_2\},\ldots\right) + 
	c\left(X_3,\ldots\right)\nonumber\\
&\text{therefore, } c\left(\{X_1,X_2,\pmb{X}',\pmb{X}''\},\ldots\right)>\nonumber\\
& c\left(X_1,\pmb{X}',\ldots\right) + c\left(X_2,\pmb{X}'',\ldots\right)
\label{eqn:superadd2}
\end{align}
where $\pmb{X}',\pmb{X}''$ are arbitrary sets of nodes in the same layer as $X_1,X_2$; potentially they could be empty sets.
\begin{lemma}
\label{lemma:rootnodeh}
Given a super-additive cost function and any pair of root nodes $\{X_1,X_2\}$ in layer $l$, the least objective function cost solution $G(\pmb{X}_L)$ has $X_1$ and $X_2$ in separate clusters.
\end{lemma}
\begin{proof}
The layer costs at layers $l'>l$ are unaffected by the cluster mapping of $X_1,X_2$ as they do not have any parent nodes by definition.
Therefore, by equation (\ref{eqn:superadd2}), any branch that contains both $X_1$ and $X_2$ in the same cluster (left-hand side) must have an objective function cost $G(\pmb{X}_L)$ greater than a branch with $X_1$ and $X_2$ in separate clusters (right-hand side), completing the proof. 
\end{proof}
Hence, for super-additive cost functions, splitting root nodes in the same layer across clusters always produces a lower cost solution. This can be used to filter cluster-layer mapping proposals as a heuristic.

\section{Appendix: Simple Example}
When run to algorithm termination (i.e.~not possible to find a better solution), DCMAP takes 315 iterations to find all three optimal solutions with total cost of 54.0 and cluster mappings of:
\begin{align*}
\renewcommand*{\arraystretch}{0.9}
k = \begin{bmatrix}
	A & B & C & D & E & F & G\\
	2 & 6 & 7 & 2 & 3 & 1 & 2\\
	1 & 6 & 7 & 2 & 3 & 1 & 2\\
	5 & 6 & 7 & 2 & 3 & 1 & 2\\
\end{bmatrix}.
\end{align*}
Note that all three optimal solutions place nodes $D$ and $G$ in the same cluster (cluster 2, shaded black in Fig. \ref{fig:eg_a}). $D$ is not only the node with the highest in-degree, but also connects the top part of the graph with the middle and, via node $G$, the bottom part of the graph. Thus it is feasible for $D$ and $G$ to be in the same cluster. Nodes $B$, $C$ and $E$ (shades of gray in Fig. \ref{fig:eg_a}) are all in individual clusters, corresponding to root ($B$, $C$) nodes or a node along a linear chain ($E$). As the objective function is based on computation cost, it is expected that root nodes are best assigned to their own clusters (Lemma \ref{lemma:rootnodeh}). The key variation between the three least-cost cluster mappings is the mapping of node $A$ which can either be part of cluster 2 with $D$ and $G$, part of cluster 1 with node $F$, or in an independent cluster of its own. 
These solutions were discovered at iterations 35, 182, and 276 on branches 21, 27 and 104, respectively. With a cost of 54.0, this is much less than the naive heuristic solution cost of 85.8.

However, near-optimal solutions were returned at earlier iterations: 
\begin{align*}
\renewcommand*{\arraystretch}{0.9}
k = \begin{bmatrix}
	A & B & C & D & E & F & G\\
	1 & 6 & 2 & 1 & 2 & 1 & 2\\
	1 & 6 & 7 & 1 & 2 & 1 & 2\\
	2 & 6 & 7 & 2 & 2 & 1  &2
\end{bmatrix}.
\end{align*}
A near-optimal solution that was within 6.7\% of the final optimal cost was discovered at iteration 9 and had a cost of 57.6.
This was refined to a cost of 57.0 at iterations 14 and 21, suggesting that the algorithm was frequently returning improved cluster mapping solutions. 

Although the earlier solutions are quite different to the optimal solution, the solution at iteration 21 starts to follow the trend of the optimal solution with $D$, $G$ and $A$ in the same cluster, with other nodes mostly in their own clusters.
This suggests that finding near-optimal or even the first optimal solution could potentially be very rapid. As might be expected, however, finding all optimal solutions could be expensive \citep{Likhachev2005,Belfodil2019}. In addition, near-optimal solutions could produce diverse cluster mappings, although they converge towards optimal mapping patterns fairly rapidly.

\section{Reproducibility Checklist for JAIR}

Select the answers that apply to your research -- one per item. 

\subsection*{All articles:}

\begin{enumerate}
    \item All claims investigated in this work are clearly stated. 
    [yes]
    \item Clear explanations are given how the work reported substantiates the claims. 
    [yes]
    \item Limitations or technical assumptions are stated clearly and explicitly. 
    [yes]
    \item Conceptual outlines and/or pseudo-code descriptions of the AI methods introduced in this work are provided, and important implementation details are discussed. 
    [yes]
    \item 
    Motivation is provided for all design choices, including algorithms, implementation choices, parameters, data sets and experimental protocols beyond metrics.
    [yes]
\end{enumerate}

\subsection*{Articles containing theoretical contributions:}
Does this paper make theoretical contributions? 
[yes] 

If yes, please complete the list below.

\begin{enumerate}
    \item All assumptions and restrictions are stated clearly and formally. 
    [yes]
    \item All novel claims are stated formally (e.g., in theorem statements). 
    [yes]
    \item Proofs of all non-trivial claims are provided in sufficient detail to permit verification by readers with a reasonable degree of expertise (e.g., that expected from a PhD candidate in the same area of AI). [yes]
    \item
    Complex formalism, such as definitions or proofs, is motivated and explained clearly.
    [yes]
    \item 
    The use of mathematical notation and formalism serves the purpose of enhancing clarity and precision; gratuitous use of mathematical formalism (i.e., use that does not enhance clarity or precision) is avoided.
    [yes]
    \item 
    Appropriate citations are given for all non-trivial theoretical tools and techniques. 
    [yes]
\end{enumerate}

\subsection*{Articles reporting on computational experiments:}
Does this paper include computational experiments? [yes]

If yes, please complete the list below.
\begin{enumerate}
    \item 
    All source code required for conducting experiments is included in an online appendix 
    or will be made publicly available upon publication of the paper.
    The online appendix follows best practices for source code readability and documentation as well as for long-term accessibility.
    [yes]
    \item The source code comes with a license that
    allows free usage for reproducibility purposes.
    [yes]
    \item The source code comes with a license that
    allows free usage for research purposes in general.
    [yes]
    \item 
    Raw, unaggregated data from all experiments is included in an online appendix 
    or will be made publicly available upon publication of the paper.
    The online appendix follows best practices for long-term accessibility.
    [yes]
    \item The unaggregated data comes with a license that
    allows free usage for reproducibility purposes.
    [yes]
    \item The unaggregated data comes with a license that
    allows free usage for research purposes in general.
    [yes]
    \item If an algorithm depends on randomness, then the method used for generating random numbers and for setting seeds is described in a way sufficient to allow replication of results. 
    [NA]
    \item The execution environment for experiments, the computing infrastructure (hardware and software) used for running them, is described, including GPU/CPU makes and models; amount of memory (cache and RAM); make and version of operating system; names and versions of relevant software libraries and frameworks. 
    [yes]
    \item 
    The evaluation metrics used in experiments are clearly explained and their choice is explicitly motivated. 
    [yes]
    \item 
    The number of algorithm runs used to compute each result is reported. 
    [yes]
    \item 
    Reported results have not been ``cherry-picked'' by silently ignoring unsuccessful or unsatisfactory experiments. 
    [yes]
    \item 
    Analysis of results goes beyond single-dimensional summaries of performance (e.g., average, median) to include measures of variation, confidence, or other distributional information. 
    [yes]
    \item 
    All (hyper-) parameter settings for 
    the algorithms/methods used in experiments have been reported, along with the rationale or method for determining them. 
    [yes]
    \item 
    The number and range of (hyper-) parameter settings explored prior to conducting final experiments have been indicated, along with the effort spent on (hyper-) parameter optimisation. 
    [yes/]
    \item 
    Appropriately chosen statistical hypothesis tests are used to establish statistical significance
    in the presence of noise effects.
    [NA]
\end{enumerate}

\subsection*{Articles using data sets:}
Does this work rely on one or more data sets (possibly obtained from a benchmark generator or similar software artifact)? 
[no]

If yes, please complete the list below.
\begin{enumerate}
    \item 
    All newly introduced data sets 
    are included in an online appendix 
    or will be made publicly available upon publication of the paper.
    The online appendix follows best practices for long-term accessibility with a license
    that allows free usage for research purposes.
    [yes/partially/no/NA]
    \item The newly introduced data set comes with a license that
    allows free usage for reproducibility purposes.
    [yes/partially/no]
    \item The newly introduced data set comes with a license that
    allows free usage for research purposes in general.
    [yes/partially/no]
    \item All data sets drawn from the literature or other public sources (potentially including authors' own previously published work) are accompanied by appropriate citations.
    [yes/no/NA]
    \item All data sets drawn from the existing literature (potentially including authors’ own previously published work) are publicly available. [yes/partially/no/NA]
    \item All new data sets and data sets that are not publicly available are described in detail, including relevant statistics, the data collection process and annotation process if relevant.
    [yes/partially/no/NA]
    \item 
    All methods used for preprocessing, augmenting, batching or splitting data sets (e.g., in the context of hold-out or cross-validation)
    are described in detail. [yes/partially/no/NA]
\end{enumerate}

\subsection*{Explanations on any of the answers above (optional):}
The cost function used in the case study only depends on the parent-child relationships in the directed acyclic graph, and the number of states of each discrete node. Hence, the only data for reproducibility is based on that to reproduce the case study examples.

The results are independent of the machine running the algorithm. Nevertheless, we used a Windows 11 PC running 14900HX CPU with 32GB of RAM and using the Armadillo library version 12.8.4. 

\end{document}